# Phantom crossing with quintom models

L. W. K. Goh [1,2]★ and A. N. Taylor[1,2]

[1] *Institute for Astronomy, University of Edinburgh, Royal Observatory, Blackford Hill, Edinburgh EH9 3HJ, UK*
[2] *Higgs Centre for Theoretical Physics, School of Physics and Astronomy, The University of Edinburgh, Edinburgh EH9 3FD, UK*

## ABSTRACT

We develop a two-scalar field quintom model, which utilizes both a quintessence-like and a phantom-like scalar field, enabling a smooth and stable transition across the $w = -1$ phantom divide as hinted by recent measurements of Baryonic Acoustic Oscillations by the Dark Energy Spectroscopic Instrument (DESI) Data Release 2. We explore a range of initial conditions and potential configurations that facilitate such a phantom-to-quintessence-like crossing, and find that this can be naturally realized with hill-top or cliff-face potentials bound from above. We study how varying these conditions affects the dynamics of the system, calculate the background observables, and compare them with DESI, CMB, and Type Ia supernova data, identifying a viable parameter space for our model. In particular, we find that a potential featuring a hyperbolic tangent form can successfully reproduce the desired phantom crossing, although such models can suffer from fine-tuning effects. Finally, we discuss prospects for distinguishing such models with upcoming state-of-the-art cosmological observations.

## 1 INTRODUCTION

The concordance ΛCDM (cold dark matter) model has been widely successful in providing an accurate description to date of the expansion history and growth of structure of our Universe. Yet a surge of high-precision observational data in recent decades has started to hint at the possibility of new physics beyond the standard model. Most recently, Baryonic Acoustic Oscillation (BAO) measurements taken by the Dark Energy Spectroscopic Instrument (DESI) Data Release 2 (DESI Collaboration 2025) show a $2.8 - 4.2\sigma$ preference for a dynamical dark energy model, whereby the dark energy equation-of-state (EoS) parameter $w$ varies with redshift, instead of being a constant $w = -1$ in the cosmological constant model. Adopting the widely used Chevallier–Polarski–Linder (CPL; M. Chevallier & D. Polarski 2001; E. V. Linder 2003) parametrization for $w$, given by

$$w(a) = w_0 + w_a(1 - a), \quad (1)$$

where $a$ is the scale factor, they find best-fitting values for $w_0$ and $w_a$ that suggest a $w < -1$ to $w > -1$ crossing going from early to late times. This crossing of the so-called 'phantom divide line' at $w = -1$ has significant theoretical implications, prompting renewed interest in models beyond ΛCDM that can naturally accommodate such behaviour (see for example G. Ye & Y.-S. Piao 2020; A. Chudaykin & M. Kunz 2024; C. Escamilla-Rivera & R. Sandoval-Orozco 2024; M. Cortês & A. R. Liddle 2024; W. Yin 2024; A. Paliathanasis 2025; H. Wang, Z.-Y. Peng & Y.-S. Piao 2025; J. Pan & G. Ye 2025; J.-Q. Wang et al. 2025; K. Lodha et al.2025 ; S. Nojiri 2025, S. D. Odintsov & V. K. Oikonomou 2025; S. Pan et al. 2025; S. Tsujikawa 2025; W. J. Wolf et al. 2025; Y. Yang et al. 2025a). In particular, scalar field models have emerged as a viable alternative explanation for dark energy.

Scalar fields are widely occurring physical constructs that have been employed in other aspects of cosmology, such as inflation theories, as well as in particle physics. When applied to dark energy, a scalar field can drive late-time cosmic acceleration through its kinetic and potential energy contributions (B. Ratra & P. J. E. Peebles 1988; C. Wetterich 1988). Depending on the dynamics, the field can exhibit an evolving EoS. Models with $w \geq -1$ are termed 'quintessence', while those with $w < -1$ are referred to as 'phantom'.

We focus on a specific class of scalar field models, dubbed the 'quintom' model (B. Feng, X. Wang & X. Zhang 2005; Z.-K. Guo et al. 2005). In such a model, phantom crossing is facilitated by the interplay between two fields, one quintessence-like and the other phantom-like, while remaining gravitationally and perturbatively stable. It was first proposed two decades ago to explain the apparent preference for $w < -1$ by Type 1a supernovae (SNe1a) data at that time (D. Huterer & A. Cooray 2005). Now, with DESI BAO data, it has once again witnessed a resurgence in interest (Y. Yang et al. 2024, 2025b; Y. Cai et al. 2025). Yet, Z.-K. Guo et al. (2006) found that when employing commonly used forms for the potential, such as an exponential or power law, quintom models that give a quintessence-to-phantom transition are more naturally realized than their phantom-to-quintessence counterparts.

In this work, we aim to develop a physically motivated quintom model that can instead naturally achieve a phantom-to-quintessence transition without introducing higher order terms, and systematically explore the sensitivity of this behaviour to variations in initial conditions and potential parameters. We then assess whether such models can be observationally constrained and distinguished from the standard ΛCDM paradigm. We shall also evaluate its feasibility by comparing the derived physical quantities with data.

★ E-mail: lgoh@roe.ac.uk

This paper has been arranged as follows: in Section 2 we give an overview of cosmological scalar field models and briefly demonstrate how single-field models cannot achieve a phantom crossing. We proceed to review the quintom model in Section 3, introducing the relevant background and perturbation equations, and explain the specific setup of our quintom model in Section 4. We then compare our results to the DESI DR2 BAO data in Section 5, and discuss potential ways to constrain such models in Section 6. Finally, we present our conclusions in Section 7.

## 2 PHANTOM CROSSING WITH SCALAR FIELDS

In the following subsection, we shall review some of the simplest examples of cosmological scalar field models, namely the quintessence model, the phantom model, and $k$-essence model, and demonstrate how such single-field models are unable to realize phantom crossing without invoking singularities or perturbation instabilities.

### 2.1 Scalar field models

The Lagrangian of a single-field quintessence model is given by

$$\mathcal{L}_\phi = X_\phi - V(\phi)\,, \tag{2}$$

with $X_\phi \equiv \frac{1}{2}\partial_\mu\phi\partial^\mu\phi = \frac{1}{2}\dot{\phi}^2$ being the kinetic energy of the canonical scalar field $\phi$ (assuming no spatial gradients) and $V$ being its potential, with a dot denoting a derivative with respect to cosmic time, $t$.

The pressure and energy density of the field are then

$$p_\phi = \frac{1}{2}\dot{\phi}^2 - V(\phi)\,, \tag{3}$$

$$\rho_\phi = \frac{1}{2}\dot{\phi}^2 + V(\phi)\,, \tag{4}$$

giving the EoS parameter

$$w_\phi \equiv \frac{p_\phi}{\rho_\phi} = \frac{X_\phi - V(\phi)}{X_\phi + V(\phi)} = \frac{\epsilon_\phi - 1}{\epsilon_\phi + 1}\,, \tag{5}$$

where we have conveniently introduced the slow-roll parameter, $\epsilon_\phi = X_\phi/V$. It is easy to see that $w_\phi \geq -1$ for all values of $\epsilon_\phi$, and that during the slow-roll regime where $V \gg X_\phi$, $\epsilon \ll 1$, and thus $w_\phi$ would tend to $-1$.

On the other hand, a phantom model is characterized by a negative kinetic energy term in the Lagrangian (R. R. Caldwell 2002)

$$\mathcal{L}_\psi = -X_\psi - V(\psi)\,. \tag{6}$$

Denoting the phantom field as $\psi$, it can also be seen that

$$w_\psi = \frac{-X_\psi - V(\psi)}{-X_\psi + V(\psi)} = \frac{-(\epsilon_\psi + 1)}{-(\epsilon_\psi - 1)}\,, \tag{7}$$

and thus $w_\psi \leq -1$. Indeed, the expression for the EoS in equation (5) is the inverse of equation (7), such that we have the duality $w \to 1/w$ when we change from simple quintessence to phantom models. Such models necessarily violate the null energy condition (NEC) of $\rho + p \geq 0$, and also potentially suffer from ghost instabilities both at the classical and quantum level, where the energy is unbounded from below (S. M. Carroll, M. Hoffman & M. Trodden 2003; J. M. Cline, S. Jeon & G. D. Moore 2004). However, extensive efforts have been put into developing models that circumvent this problem; see for example A. E. Schulz & M. White (2001); B. McInnes (2002); V. K. Onemli & R. P. Woodard (2002, 2004); M. P. Dabrowski, T. Stachowiak & M. Szydłowski (2003); P. H. Frampton (2003); V. Sahni & Y. Shtanov (2003); P. Singh, M. Sami & N. Dadhich (2003); I. Y. Aref'eva, S. Y. Vernov & A. S. Koshelev (2006).

From equations (2) and (6), it is easy to see that the EoS of quintessence and phantom models with Lagrangians of this form will always remain on either side of the $w = -1$ phantom boundary. Allowing $w$ to cross this boundary would require introducing higher order terms in the Lagrangian, such as those of $k$-essence models (T. Chiba, T. Okabe & M. Yamaguchi 2000; C. Armendariz-Picon, V. Mukhanov & P. J. Steinhardt 2000, 2001). Specifically, $k$-essence models admit a kinetic term that is a generalized function of $X$, and can realize either a $w < -1$ or $w > -1$ scenario while being stable to ghosts. Its Lagrangian is given by

$$\mathcal{L}_k = F(X) - V\,, \tag{8}$$

generating a more complex form for the EoS

$$w_k = \frac{F - V}{2XF_X - F + V}\,, \tag{9}$$

where $F_X = \partial F/\partial X$. Consequently, $w_k$ can reside on either side of the phantom boundary depending on the choice of $F$ and $V$.

### 2.2 Phantom crossing No-Go theorem

On the other hand, A. Vikman (2005) demonstrated that generally, there exists a 'No-Go Theorem' whereby phantom crossing cannot be fulfilled with a single scalar field model, as this necessarily induces perturbation instabilities. We shall illustrate this with a simple example for the $k$-essence model.

We define the effective adiabatic sound speed of the scalar field as

$$c_s^2 \equiv \frac{p_X}{\rho_X}\,, \tag{10}$$

with the condition that $c_s^2 \geq 0$. Here, the $X$ subscript denotes a partial derivative with respect to $X$. From equation (9) this gives

$$c_s^2 = \frac{F_X}{2XF_{XX} + F_X}\,. \tag{11}$$

At the point of crossing $w = -1$, $F_X = 0$, and $c_s^2 = 0$; when $w < -1$, $F_X < 0$, and thus $c_s^2 < 0$, resulting in unphysical gradient instabilities.

R. R. Caldwell & M. Doran (2005) and W. Hu (2005) then showed that in order to achieve phantom crossing while remaining stable both at the background and perturbation levels, the scalar field requires at least one extra degree of freedom. This motivates our exploration of a two-field scalar model instead, which we will discuss in detail for the rest of this work.

## 3 THE QUINTOM MODEL

The quintom model was first proposed by B. Feng et al. (2005), Z.-K. Guo et al. (2005), and postulates the simultaneous existence of two scalar fields: a quintessence-like $\phi$ field and a phantom-like $\psi$ field, which can easily realize the $w = -1$ crossing while adhering to the stability conditions introduced in the previous section. While the two-scalar field model is perhaps the simplest model to satisfy our constraints for stable phantom crossing, we can still regard it as an effective field theory of some more fundamental approach, which captures the extra degree of freedom. In this section, we shall explain the dynamics of such a model, detailing the background and perturbation equations, and discuss a specific setup we will adopt to study its feasibility in realizing a $w$ crossing as hinted by DESI data.

### 3.1 Background equations

In a quintom model, instead of a single scalar field, the dark energy sector is comprised of two fields, which are assumed to be minimally coupled, with the Lagrangian

$$\mathcal{L} = X_\phi - X_\psi + V(\phi) + V(\psi). \tag{12}$$

We can see how such differential kinetic terms may arise if we consider the two scalar fields as directions in a field space. We can then rotate the fields to form two orthogonal fields, $\Theta = \phi + \psi$ and its complement $\bar{\Theta} = \phi - \psi$. We can form a kinetic energy term for this rotated field in the Lagrangian as $\dot{\Theta}\dot{\bar{\Theta}}/2 = X_\phi - X_\psi$. This can become negative, breaking the NEC, but we have the condition that the total energy density, $\rho_\Theta = \rho_\phi + \rho_\psi$, remains positive. The fields then separately obey the Klein–Gordon equation,

$$\ddot{\phi} + 3H\dot{\phi} + \frac{\partial V(\phi)}{\partial \phi} = 0\,, \tag{13}$$

$$\ddot{\psi} + 3H\dot{\psi} - \frac{\partial V(\psi)}{\partial \psi} = 0\,. \tag{14}$$

Here $H \equiv \frac{\dot{a}}{a}$ is the Hubble parameter and $V(\phi)$ and $V(\psi)$ are the potentials of each field. Note that since $\psi$ is a phantom-like field with a negative kinetic term in the Lagrangian, the sign in front of the derivative of the potential is negative in equation (14). Thus, we can appreciate that the phantom field will run *up* the potential, with an increasing energy density $\rho_\psi$. Conceptually, it is useful to note that the Klein–Gordon equation can be interpreted as an 'inversion' of the potential, so that our usual intuition about motion in a potential field can be applied.

Assuming a flat Friedmann–Lemaître–Robertson–Walker (FLRW) metric, the conservation equation then reads as

$$H^2 = \frac{8\pi}{3}(\rho_\mathrm{m} + \rho_\mathrm{r} + \rho_\phi + \rho_\psi), \tag{15}$$

where we have taken $\hbar = G = c = 1$, while the subscript 'm' denotes non-relativistic matter (baryons and cold dark matter), 'r' denotes radiation, and the energy densities of the fields are defined as $\rho_\phi = \frac{1}{2}\dot{\phi} + V(\phi)$ and $\rho_\psi = -\frac{1}{2}\dot{\psi} + V(\psi)$, which make up the dark energy sector.

We can then consider the EoS of each field, $w_\phi$ and $w_\psi$, given by

$$w_\phi = \frac{p_\phi}{\rho_\phi} = \frac{\frac{1}{2}\dot{\phi}^2 - V(\phi)}{\frac{1}{2}\dot{\phi}^2 + V(\phi)}\,, \tag{16}$$

$$w_\psi = \frac{p_\psi}{\rho_\psi} = \frac{-\frac{1}{2}\dot{\psi}^2 - V(\psi)}{-\frac{1}{2}\dot{\psi}^2 + V(\psi)}\,, \tag{17}$$

which must separately remain within their respective $w = -1$ boundary ($w_\phi > -1$ for the $\phi$ field and $w_\psi < -1$ for the $\psi$ field). However, the total effective dark energy EoS of the combined fields,

$$w_\mathrm{DE} = \frac{p_\phi + p_\psi}{\rho_\phi + \rho_\psi} = \frac{\frac{1}{2}\dot{\phi}^2 - \frac{1}{2}\dot{\psi}^2 - V(\phi) - V(\psi)}{\frac{1}{2}\dot{\phi}^2 - \frac{1}{2}\dot{\psi}^2 + V(\phi) + V(\psi)}\,, \tag{18}$$

can transition across this boundary, hence achieving the desired crossing behaviour. This is further elucidated by once again recasting $w_\mathrm{DE}$ in terms of the slow-roll parameter $\epsilon_\mathrm{quintom}$, where this time

$$\epsilon_\mathrm{quintom} \equiv \frac{X_\mathrm{quintom}}{V_\mathrm{quintom}} = \frac{\frac{1}{2}(\dot{\phi}^2 - \dot{\psi}^2)}{V(\phi) + V(\psi)}\,. \tag{19}$$

In the slow-roll regime where $\epsilon_\mathrm{quintom} \ll 1$,

$$w_\mathrm{DE} = \frac{X_\mathrm{quintom} - V_\mathrm{quintom}}{X_\mathrm{quintom} + V_\mathrm{quintom}} = \frac{\epsilon_\mathrm{quintom} - 1}{\epsilon_\mathrm{quintom} + 1} \approx -1 + 2\epsilon_\mathrm{quintom}\,. \tag{20}$$

This has the same form as the EoS for the quintessence field, but the slow-roll parameter can now become negative, allowing the combined field to track both quintessence and phantom behaviour. The crossing of $w_\mathrm{DE}$ is solely dictated by the difference in speeds of the fields: when the phantom field is moving faster, the system is in a phantom regime where $w_\mathrm{DE} < -1$, and when the quintessence field is dominating, we transition into $w_\mathrm{DE} > -1$. This gives us an easier handle with which to control the dynamics of the system, whereby it simply boils down to controlling the speeds of the fields.

### 3.2 Linear perturbation equations

We can now turn to the modified perturbation equations of a quintom model, since we want to assess if the model is stable at the level of perturbations. We consider the perturbed metric in the conformal Newtonian gauge,

$$\mathrm{d}s^2 = a(t)^2\left[(1 + 2\Psi)\,\mathrm{d}t^2 - (1 - 2\Phi)\,\mathrm{d}x^i\mathrm{d}x_i\right], \tag{21}$$

where $\Psi$ is the Newtonian gravitational potential, $\Phi$ the spatial curvature potential, while assuming $\Psi = \Phi$ due to the absence of anisotropic stress. Defining the overdensity as $\delta_i = (\rho_i - \bar{\rho}_i)/\bar{\rho}_i$ for $i = \{\phi, \psi\}$ and velocity divergence $\theta_i = \frac{ik^j\delta T^0_j}{\rho_i + p_i}$, we can write down the evolutions of the scalar field overdensity and velocity divergence (C.-P. Ma & E. Bertschinger 1995)

$$\dot{\delta}_i = -(1 + w_i)(\theta_i - 3\dot{\Phi}) - 3H\left(\frac{\delta p_i}{\delta \rho_i} - w_i\right)\delta_i\,, \tag{22}$$

$$\dot{\theta}_i = -H(1 - 3w_i)\,\theta_i - \frac{\dot{w}_i}{1 + w_i}\theta_i + \frac{\delta p_i}{\delta \rho_i}\frac{1}{1 + w_i}k^2\delta_i + k^2\Psi\,, \tag{23}$$

where $k$ is the Fourier space wavenumber. Since we have assumed minimal coupling between the fields, they are presumed to obey these evolution equations separately without inducing a mixing of terms. It is also easy to appreciate from equation (23) that at the crossing point of $w_i = -1$, the denominator $(1 + w_i)$ in the second and third terms will diverge, once again demonstrating that single-field scalar models are perturbatively unstable at phantom crossing.

On the other hand, the effective dark energy overdensity and velocity divergence of a two-field quintom system is the effective sum of that of each field (G.-B. Zhao et al. 2005)

$$\delta = \frac{\Sigma_i\,\rho_i\,\delta_i}{\Sigma\,\rho_i}\,, \tag{24}$$

$$\theta = \frac{\Sigma_i\,(\rho_i + p_i)\,\theta_i}{\Sigma_i\,(\rho_i + p_i)}\,. \tag{25}$$

Following the definition of the sound speed squared $c^2_{s,i} \equiv \frac{\delta p_i}{\delta \rho_i}$ from the previous section, we can see that in the quintom model, $c^2_{s,i} = 1$ for both the quintessence and phantom field due to the canonical kinetic term in the Lagrangian, hence overcoming the issue of gradient instabilities as well.

G.-B. Zhao et al. (2005) provide a full derivation of the adiabatic and isocurvature modes of the scalar field perturbations in a quintom model. They also demonstrate that for the regions where $w < -1$ or $w > -1$, the perturbations of a quintom model behave as a single field phantom and quintessence model, respectively, while in the phantom crossing region where $-1 - c < w < -1 + c$ for values of $c < 10^{-5}$, $\dot{\delta}_\mathrm{q}$ and $\dot{\theta}_\mathrm{q}$ can be approximated to 0. W. Hu & I. Sawicki (2007) and W. Fang, W. Hu & A. Lewis (2008) have also developed the parametrized post-Friedmann (PPF) approach to calculate dark energy perturbations at the $w = -1$ boundary for dark energy models. Hence we find the quintom model is stable at the level of perturbations. Given this, we now study the dynamics of the background.

## 4 MODEL SETUP

We shall now look into the background dynamics of a quintom model by specifying characteristic forms for the scalar field potentials $V(\phi)$ and $V(\psi)$, and numerically solving equations (13), (14), and (15) to derive the expansion history and EoS parameters. Z.-K. Guo et al. (2006) demonstrated that quintom models can be broadly classified into two types: a quintom-A type where the quintessence field dominates at early times before transitioning to a phantom-dominated one at late times (the dark energy EoS transitions from $w_{\rm DE} > -1$ to $w_{\rm DE} < -1$), and a quintom-B type where the phantom field dominates before transitioning to a quintessence regime ($w_{\rm DE} < -1$ to $w_{\rm DE} > -1$). Here, we shall particularly look at the latter class of quintom models, which has been shown to be preferred by the data. Past literature demonstrated that quintom-A type models are more readily achieved with simple forms for the potential and natural attractor solutions, while quintom-B type models necessitate a certain degree of fine-tuning (Z.-K. Guo et al. 2006; Y.-F. Cai et al. 2010; Y. Yang et al. 2025b). We can understand this behaviour by first considering inverse power-law potentials, such as $V(\phi) \propto \phi^{-\alpha}$ and $V(\psi) \propto \psi^{-\alpha}$, with positive index $\alpha$. If initially the quintessence field, $\phi$, starts higher up than the phantom field, $\psi$, each with low initial velocity, the quintessence field will roll down faster and have higher potential energy, dominating over the phantom field. However, the phantom field will roll up the potential, gaining momentum until its kinetic and potential energy dominate that of the quintessence field. Hence, the system will generically tend to favour a quintessence-to-phantom behaviour. We can see this more clearly if we assume the slow-roll approximation, where $\dot{\phi} \approx -V'(\phi)/3H$ and $\dot{\psi} \approx V'(\psi)/3H$, where $V'$ is the gradient of the potential, so that

$$\epsilon_{\rm quintom} \approx \frac{|V'(\phi)|^2 - |V'(\psi)|^2}{48\pi\,(V(\phi) + V(\psi))^2}, \tag{26}$$

where $\phi$ starts on a steeper gradient than $\psi$. Exponential potentials will give similar behaviour.

In order to realize a phantom-to-quintessence (i.e. quintom-B type) crossing, the speed of the phantom field $\dot{\psi}$ should initially dominate over that of the quintessence field $\dot{\phi}$ to have $w_{\rm DE} < -1$, with $\dot{\phi}$ then dominating only at late times for the $w_{\rm DE} > -1$ transition to occur. I. Y. Aref'eva, A. S. Koshelev & S. Y. Vernov (2005), S. Y. Vernov (2008), A. Gómez-Valent & A. González-Fuentes (2025) have similarly considered viable two-field models within additional setups of, respectively, a brane approach, string theory or negative quintessence with multiple potentials. However, in this work we shall aim to develop the simplest model with the least degrees of freedom, hence we restrict $\phi$ and $\psi$ to the same form of the potential, i.e. having $V(\phi) = V(\psi)$.

We can imagine that crossing can be achieved by having the quintessence field start with a smaller slope [i.e. smaller $V'(\phi)$], and hence velocity, $\dot{\phi}$. We can do this by bounding the potential from *above* with a hill-top potential, such that the quintessence field is still initially higher than the phantom field, but with a lower velocity. If the phantom field is on the slopes of the potential, it will quickly pick up speed, dominating over the quintessence kinetic energy. When the quintessence field starts to fall, it will gain speed, while the phantom field will reach the peak of the hill-top potential. There it will oscillate around the peak before coming to a halt at the peak. Meanwhile, the quintessence field will continue to roll down the slopes of the potential and become the dominant kinetic energy. Although there is some fine-tuning in the initial conditions, hill-top models will generically give rise to quintessence-to-phantom behaviour. We shall investigate different forms for $V(\phi)$ and $V(\psi)$, and derive constraints on the initial conditions that could create such a scenario.

For our analysis, the parameters of the field that we will vary are $\{\phi_{\rm ini}, \psi_{\rm ini}, \dot{\phi}_{\rm ini}, \dot{\psi}_{\rm ini}\}$, where the 'ini' subscript refers to the values of the fields and their speeds at some initial redshift from which we evolve our system, which we have set to $z_{\rm ini} = 10^{20}$. To enforce a quintom-B type transition, we additionally impose the condition $|\dot{\psi}_{\rm ini}| \geq |\dot{\phi}_{\rm ini}|$ such that $w_{\rm DE,ini} \leq -1$. We shall also only consider positive values of the fields, where $\phi > 0$ and $\psi > 0$. Furthermore, we fix the value of the present-day matter density $\Omega_{\rm m,0}$ and the Hubble parameter $H_0$ to the best-fitting values obtained by the DESI DR2 BAO+CMB+Pantheon+data set for a $w_0w_a$CDM cosmology (DESI Collaboration 2025), where $\Omega_{\rm m,0} = 0.3114$ and $H_0 = 67.15\,{\rm km\,s^{-1}\,Mpc^{-1}}$, as well as fixing the present-day radiation density $\Omega_{\rm r,0} = 9.23 \times 10^{-5}$ to the *Planck* 2018 best-fitting $\Lambda$CDM cosmology (Planck Collaboration VI 2020).

### 4.1 Gaussian potential

To facilitate a quintom-B type transition, we can consider a monotonically decaying potential, for example a Gaussian function that has a similar expression to the exponential form for $V$, except with a hilltop cutoff at the peak, as described in the preceding section. We express it as

$$V(\phi) = V_0\, e^{-\frac{(\phi-\mu)^2}{2\sigma^2}}, \tag{27}$$

where we fix $\mu < \phi_{\rm ini} < \psi_{\rm ini}$. For simplicity, we shall assume a similar potential for $\psi$. We can then further vary the parameters of the potential $\{V_0, \mu, \sigma\}$, on top of those of the fields.

We numerically solve for the expansion history of this model, setting values of $\phi_{\rm ini} = 1.04\,m_{\rm P}$, $\psi_{\rm ini} = 1.32\,m_{\rm P}$, $\dot{\phi}_{\rm ini} = 1 \times 10^{-5}\,m_{\rm P}^2$, $\dot{\psi}_{\rm ini} = 1.2 \times 10^{-5}\,m_{\rm P}^2$, $V_0 = 5 \times 10^{-8}\,m_{\rm P}^4$, $\mu = 0.98\,m_{\rm P}$, $\sigma = 0.23\,m_{\rm P}$, which have been roughly calibrated to reach reasonable values of $\Omega_{\rm m,0}$, $\Omega_{\Lambda,0}$, and $H_0$. Here $m_{\rm P}$ is the reduced Planck mass, given by $m_{\rm P} = \frac{1}{\sqrt{8\pi G}}$.

In Fig. 1 we plot the characteristics of the fields: their potentials, evolution of their speeds, and the resultant energy densities. We see that at first order, we are able to achieve phantom crossing with the desired behaviour as laid out in the preceding paragraphs. The phantom field starts at the bottom of the potential, while the quintessence field is stationary and close to the peak. Up to redshifts of approximately $z > 5$, the fields are frozen until the dark energy component begins to dominate, and the phantom (quintessence) field begins rolling up (down) the potential. Both fields gain speed, sourced by the $\partial V/\partial\phi$ term in the Klein–Gordon equation, and eventually cross. The phantom field slightly speeds up before slowing down again as it reaches the peak, while the quintessence field continues accelerating, reaching a terminal velocity. This drives a momentary increase in $\delta\dot{\psi}$, before it decreases again and becomes negative.

In terms of the dark energy density, we see that as expected of individual quintessence and phantom field behaviour, the energy density increases for the phantom case, since the increase in $V(\psi)$ is larger than the increase in magnitude of $\dot{\psi}$, while the quintessence energy density decreases, as the decrease in $V(\phi)$ is larger than the increase in $\dot{\phi}$. Overall, this leads to an increase in the total dark energy density, peaking before decreasing slightly during quintessence dominance. In all cases, the energy density of the individual and combined fields remains positive.

We then present $w(z)$ for each field, as well as the effective dark energy EoS $w_{\rm DE}(z)$ in Fig. 2. We see that at early times, $w(z)$ tends to a value of $-1$. This is because, despite the non-zero initial speeds of both fields, the system is still deep within the radiation-dominated era, whereby they are essentially frozen. Thus, they can be approximated as being in a slow-roll regime where $\epsilon_{\rm quintom} \ll 1$ and therefore $w(z) \approx -1$. We then see the gradual decrease in $w_{\rm DE}(z)$ at late times, corresponding to the speed-up of the phantom field before the quintessence field dominates and $w_{\rm DE}(z) > -1$. When we plot the individual EoS parameters of each field, we see that they do not cross the $-1$ boundary, hence remaining stable to perturbations.

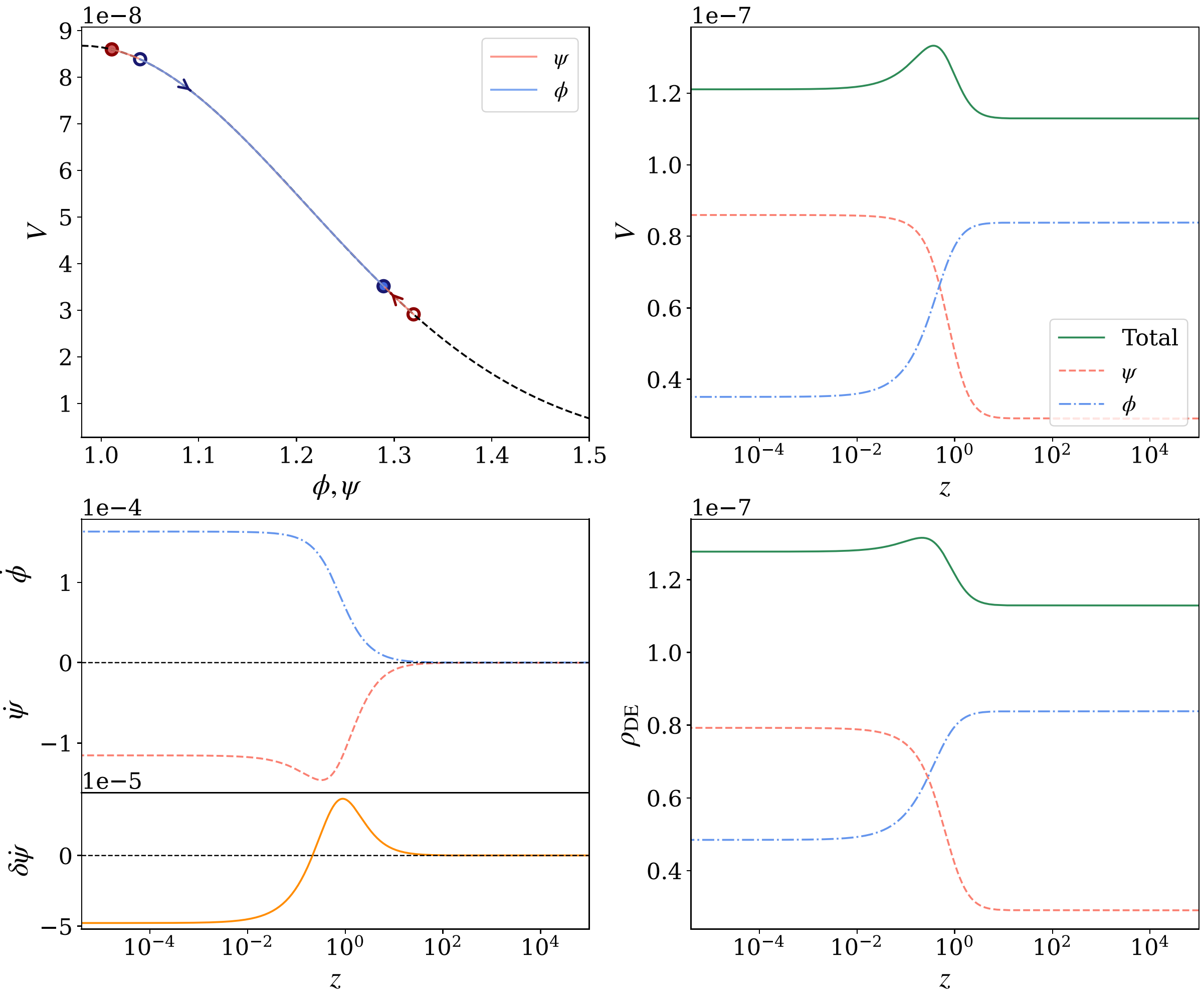

**Figure 1.** Clockwise from top left: plot of the Gaussian potential against the value of the field (for the $\psi$ field in pink and $\phi$ field in blue). The unfilled dots mark the initial point of the field on the potential, the filled dots mark the ending point (when $z = 0$), while the corresponding coloured arrows mark the direction of evolution. The Gaussian function of the field has also been plotted in dashed lines. Plot of the potential against redshift $z$, for the $\psi$ field (dashed), $\phi$ field (dashed dotted), and the total potential of the system (solid). Plot of the energy density of the dark energy sector $\rho_{\rm DE}$, for each field as well as their sum. Plot of the speeds of the fields $\dot{\psi}$ and $\dot{\phi}$, as well as their difference in magnitude $\delta\dot{\psi} = |\dot{\psi}| - |\dot{\phi}|$ in the bottom plot in solid lines. Note that here (and in the rest of the plots), we have omitted the units of the parameter values for brevity.

Subsequently, we vary the initial conditions and parameters of the model $\{\dot{\phi}_{\rm ini}, \dot{\psi}_{\rm ini}, \phi_{\rm ini}, \psi_{\rm ini}, V_0, \mu, \sigma\}$ to see how these affect the dynamics of the system. The full results are presented in Appendix A1; here we only summarize the key points. Primarily, we see that the system is insensitive to the initial values of the field speed $\dot{\phi}_{\rm ini}$ and $\dot{\psi}_{\rm ini}$ since, as we have discussed previously, the fields are frozen at early times and will always tend to a slow-roll regime with $w(z) \approx -1$. From equation (27) and the Klein–Gordon equations, we can appreciate that degeneracies exist between the rest of the parameters, since they all work to alter the values of $\partial V(\phi)/\partial\phi$ and $\partial V(\psi)/\partial\psi$ and thus the speeds of the fields, consequently changing the behaviour of $w(z)$. Furthermore, we can see a natural degeneracy between the values of $\mu$, $\phi_{\rm ini}$ and $\psi_{\rm ini}$: it is their relative differences that affect the dynamics of the system and not their absolute value.

We can gain analytical insight into how this behaviour manifests based on the expression for the derivative of the potential, given by

$$\frac{\partial V}{\partial \phi} = \frac{-V_0\,(\phi - \mu)}{\sigma^2} e^{\frac{-(\phi-\mu)^2}{2\sigma^2}} . \tag{28}$$

Generally, we can infer that the larger the magnitude of $V_0$ and $\mu$, the larger the initial value of $\partial V/\partial\phi$, which then drives the acceleration of the fields as seen from equations (13) and (14). The inverse behaviour is seen when varying $\sigma$. On the other hand, changes in $\phi_{\rm ini}$ or $\psi_{\rm ini}$ affect primarily the dynamics of each individual field. For a more in-depth analysis, we refer the reader to Appendix A1.

Interestingly, we also note that in some cases where the speed of the phantom field is sufficiently large when it reaches the peak of the potential, it might begin to ‘roll down’ the opposite side of the peak. If we consider it as the inverse of the potential, we expect that as the phantom field overshoots the peak, it will oscillate around it before settling down at the maximum point. In this case, the quintessence field will again dominate, resulting in a decrease in $w_{\rm DE}$ in the future.

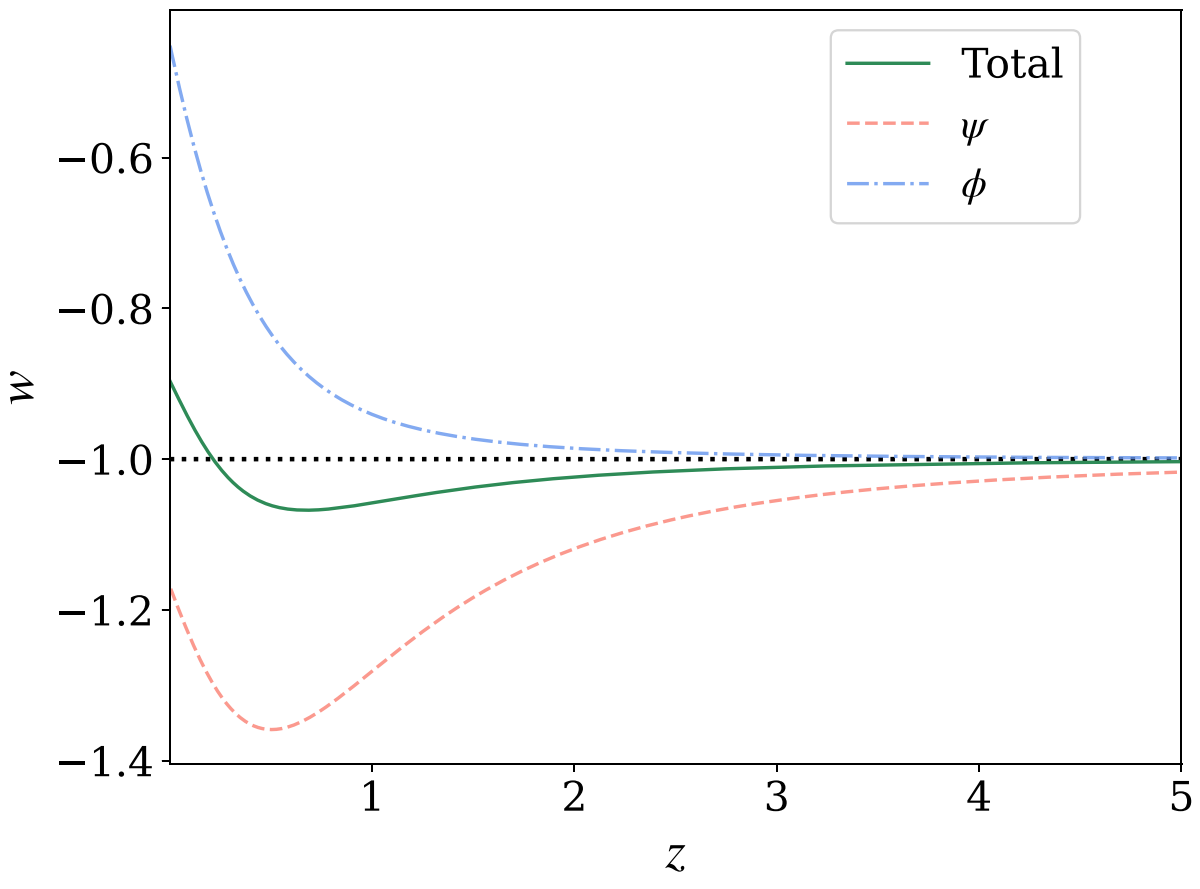

**Figure 2.** Plot of the evolution of the dark energy EoS $w_{\rm DE}(z)$ for the quintessence field (in dashed dotted), the phantom field (in dashed), and the total effective dark energy EoS (in solid), assuming a Gaussian form for the scalar field potentials. We also denote the $w = -1$ boundary in dotted grey lines.

This initial study demonstrates that quintom-B type crossings can indeed be achieved with a Gaussian potential. From Fig. A1, we see that the evolution of $w(z)$ is sensitive to the initial values of the fields $\phi_{\rm ini}$ and $\psi_{\rm ini}$, as well as the parameters of $V(\phi)$ and $V(\psi)$. Given that the current combined constraints on the present-day EoS value, $w_{\rm DE,0}$, from BAO, CMB, and SNe1a data are already at an approximately 5 per cent level, we can imagine that this would admit a fairly restricted viable parameter space. We shall investigate this further in the following sections.

### 4.2 Hyperbolic tangent potential

We additionally consider a similarly decaying potential, except with a plateau instead of a peak. This essentially freezes the quintessence field until sufficiently late times, thereby allowing for an evolution to more negative values of $w_{\rm DE}$ before the phantom transition. Correspondingly, the phantom field gets 'frozen out' when it reaches the plateau, where it would naturally decelerate and reach a steady state more rapidly than in the Gaussian potential case.

Such a form for $V(\phi)$ could be parametrically reconstructed with a step-like, or 'cliff-face', hyperbolic tangent function, where we have chosen to adopt the expression

$$V(\phi) = V_0\left(\tanh\left[s_1(s_2 - \phi)\right] + 1\right) , \qquad (29)$$

with the parameter $s_1$ controlling the slope of the potential and thus the speeds of the fields, while $s_2$ controls the value of the field where the drop in the potential occurs. However, since there is a degeneracy between the relative differences of $s_2$, $\phi$, and $\psi$ as was the case for a Gaussian potential, we can reduce the parameter space by fixing $s_2 = 1\,m_{\rm P}^{-1}$ and only allowing the values of $\phi$ and $\psi$ to vary. We shall also rename $s_1$ to $s$ hereafter for brevity. For completeness, we also write down the derivative of the potential as

$$\frac{\partial V}{\partial \phi} = -sV_0\left(1 - \tanh^2\left[s\left(1-\phi\right)\right]\right) . \qquad (30)$$

We set $\psi_{\rm ini} = 1.02\,m_{\rm P}$, $\phi_{\rm ini} = 0.92\,m_{\rm P}$, $\dot{\psi}_{\rm ini} = \dot{\phi}_{\rm ini} = 1\times10^{-5}\,m_{\rm P}^2$, $V_0 = 0.91\times10^{-8}\,m_{\rm P}^4$, $s = 29\,m_{\rm P}^{-1}$, and plot our results in Figs 3 and 4. We see that similar behaviour to that of a Gaussian potential is achieved, except that the phantom field experiences a stronger deceleration when it rolls up the plateau of the potential. In the case of the EoS parameter, the decrease in $w_{\rm DE}(z)$ occurs at earlier redshifts, mildly oscillating at around $z = 1$ due to the change in the profiles from phantom to the quintessence-dominated regime.

We vary the six parameters of the system and present the full results in Appendix A2. Generally, a phantom crossing is more easily realized in the case of a hyperbolic tangent potential than with a Gaussian function. Once again, the system is insensitive to the magnitude of the initial speeds of the fields as the fields tend to the slow-roll attractor solution at early times. Varying the initial value of the fields then dictates how far up or down the potential they evolve: the larger their value, the greater the magnitude of $\partial V/\partial\phi$ and thus the further down (or up, in the case of the phantom field) it rolls. We see that the system is highly sensitive to its initial values: for example, a shift of 0.02 per cent in $\phi_{\rm ini}$ gives an approximately 35 per cent increase in $w_{\rm DE,0}$. Similar behaviour can be seen when varying $V_0$ and $s$. Given the precision of current observational data, the viable parameter space can once again already be tightly constrained.

To investigate this in more detail, we present in Fig. 5 a 2D heatmap of $w_{\rm DE,0}$ when conducting a parameter sweep across $\phi_{\rm ini}$ and $s$, for three chosen values of $V_0$. From here, we can begin to mark out the regions of the 2D parameter space that give unphysically high values of $w_{\rm DE,0}$, which is already ruled out by most observational data sets. Taking a step further, if DESI data is presumed to be accurate, by marking out the region that gives a $w_{\rm DE,0}$ that is within $1\sigma$ of the DESI best-fit (between the white and light brown lines in the figure), we can see that this admits a very restricted area of the parameter space.

Additionally, from Fig. 5, two degeneracies clearly stand out: first, for the same $V_0$ and $s$, two values of $\phi_{\rm ini}$ can give the same $w_{\rm DE,0}$. We investigate this behaviour further by choosing two random values of $\phi_{\rm ini}$ along the axis of $s = 25.8\,m_{\rm P}^{-1}$ that give the same value of $w_{\rm DE,0} = -0.7$, and plot their $w_{\rm DE}(z)$ evolutions in Fig. 6. It is then clear to see that this degeneracy arises due to the fact that at large enough values of $\phi$ (represented by the model drawn out in the blue solid line), instead of being in a state where the field is still rolling down the potential at present-day (as depicted in Figs 3 and 4) and thus having an increasing $w_{\rm DE}(z)$, the quintessence field has hit the lower plateau of the potential and is slowing down. This manifests as a decrease in $w_{\rm DE}(z)$, after having peaked at an earlier redshift. In fact, we can extrapolate that this would be the fate of the universe should the quintom model drawn out by the orange line (and favoured by the data) be left to evolve into the future. Subsequently, $w_{\rm DE}$ would asymptote to $-1$ as both fields slow down, returning the system to a cosmological-constant-like setup. Therefore, the region of the parameter space that lie to the right of the yellow 'ridge' of $w_{\rm DE,0}$ maxima in Fig. 5 correspond to the class of models which can already be ruled out by the data, as they have evolved from a state where $w_{\rm DE}(z)$ has reached a peak in the quintessence regime at earlier times, and is decreasing from above.

The second degeneracy is the linear relationship between $s$ and $\phi_{\rm ini}$. For increasing values of $V_0$, this degeneracy line occurs at smaller values of $\phi_{\rm ini}$. To gain insight into this behaviour, we solve for the dynamics of the model for two arbitrarily chosen sets of parameter values: $\{s = 27.515, \phi_{\rm ini} = 0.931\}$ and $\{s = 22.525, \phi_{\rm ini} = 0.925\}$, that both give a value of $w_{\rm DE,0} = -0.69$. We plot the results in Fig. 7, where the resultant $w_{\rm DE}$ evolution is presented in the top plot, and the potential of the quintessence field in the bottom plot. We can see that even though both sets of parameter values give the same value of $w_{\rm DE,0}$, their evolutions differ slightly, due to how $s$ and $\phi_{\rm ini}$ simultaneously work to modify the initial value of $\partial V/\partial\phi$ and subsequently $\dot{\phi}_{\rm ini}$, which can also be inferred from equation (30).

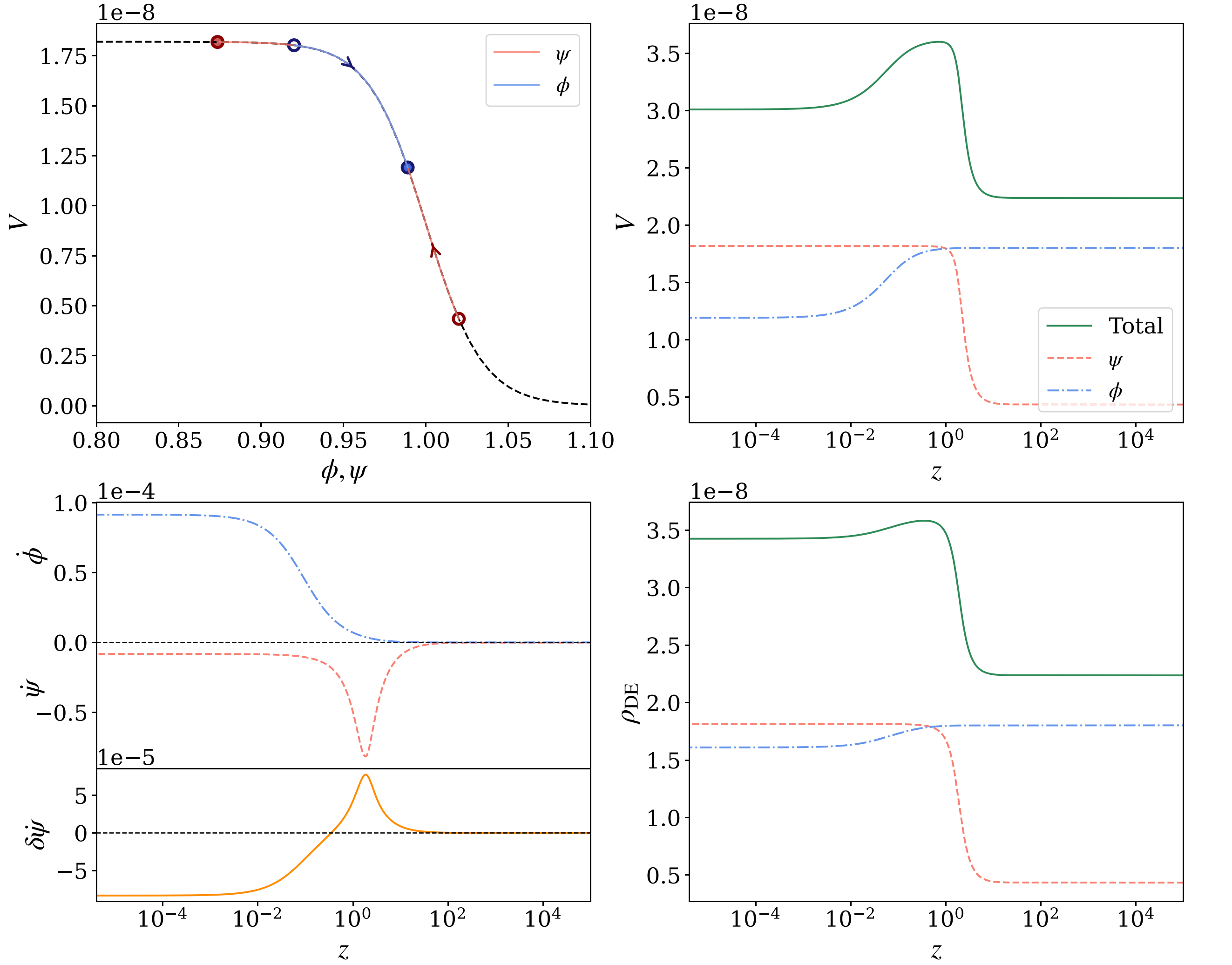

**Figure 3.** Similar to Fig. 1, for a hyperbolic tangent potential. Plot of (clockwise from top left): the potential of both fields, their evolution as a function of redshift, the dark energy density, and the speeds of both fields and their difference.

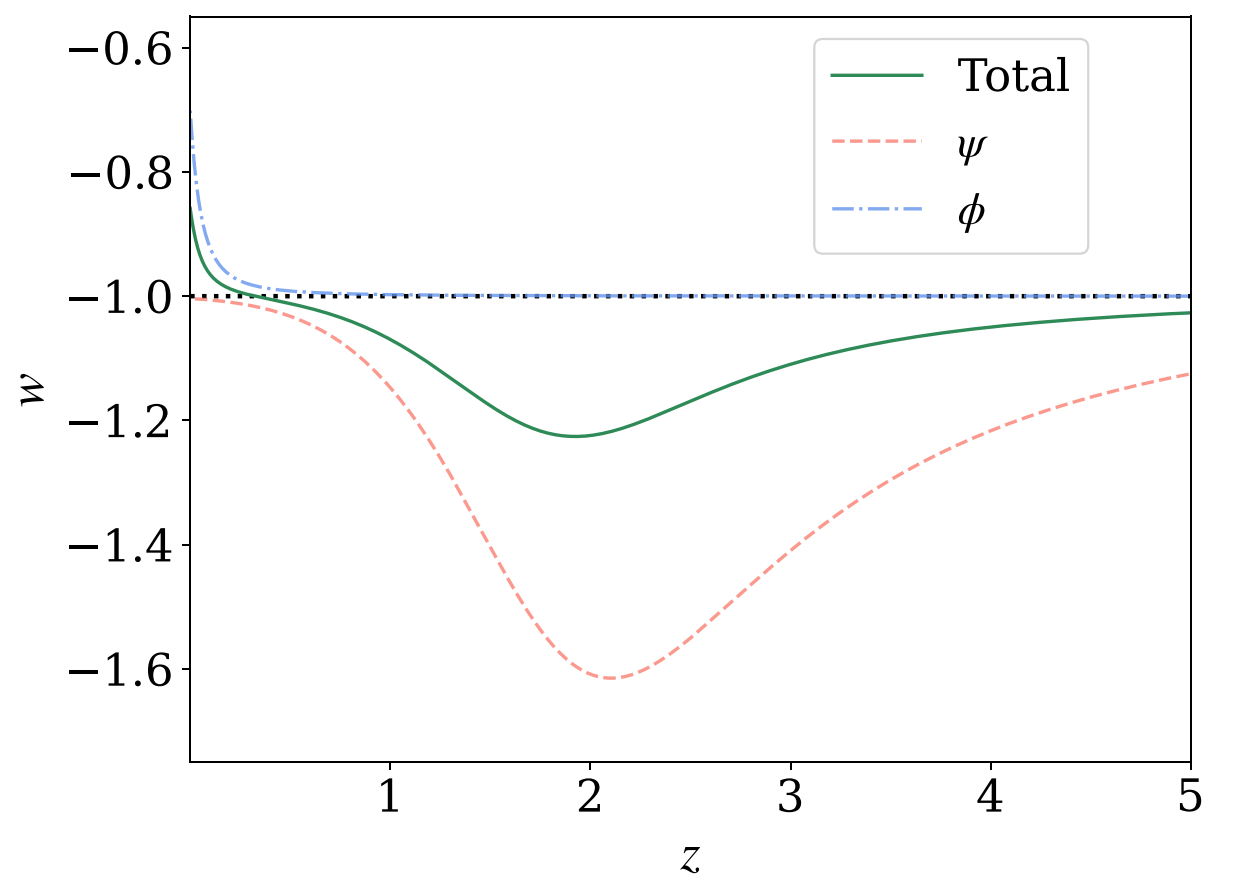

**Figure 4.** Plot of the evolution of the dark energy EoS similar to Fig. 2, for a hyperbolic tangent potential.

We also stress that these sets of parameter values (for both Figs 6 and 7) have been chosen for illustrative purposes to investigate the degenerate nature between various initial conditions, and not as a proposition for a viable models that can fit the data.

## 5 COMPARISON TO DATA

Having solved for the background dynamics of our quintom model and studied the behaviour of the system, we can assess how well it performs in fitting the data, specifically BAO distance measurements from DESI Data Release 2 (DESI Collaboration 2025). In the following analyses, we adopt the initial values specified in Section 4.2 for a hyperbolic tangent potential. However, we caution that since this model has not been rigorously fitted to the data through Bayesian inference analysis, the values of the initial conditions are by no means their best-fitting values, but merely chosen as an illustrative example and proof of concept of our model. From there, we also compare the various physical background quantities derived from a general $w_0 w_a$ CDM model against those calculated from a quintom model, to establish the extent to which a quintom model can reproduce the observed behaviour captured by the CPL parametrization.

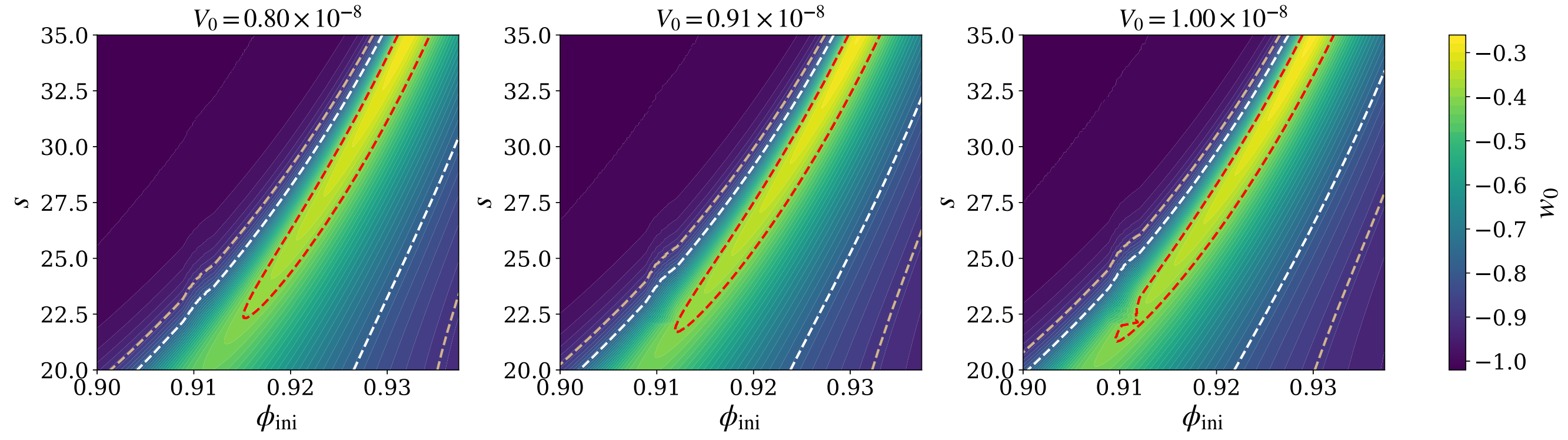

**Figure 5.** Heatmap of resultant $w_{\rm DE,0}$ values when varying the parameters $\phi_{\rm ini}$ and $s$ of the system, for values of $V_0 = \{0.80 \times 10^{-8}\, m_{\rm P}^4, 0.91 \times 10^{-8}\, m_{\rm P}^4, 1.00 \times 10^{-8}\, m_{\rm P}^4\}$. The rest of the parameters have been fixed to their values stated in the text. The red contour demarcates the region of $w_0$ that is $3\sigma$ above the best fit when employing the DESI BAO+CMB+Union3 data set (the least restrictive), while the white and light brown contours mark the region where $w_0$ is $1\sigma$ above and below the best-fit value respectively, when using the DESI BAO+CMB+Pantheon+data set (the most restrictive).

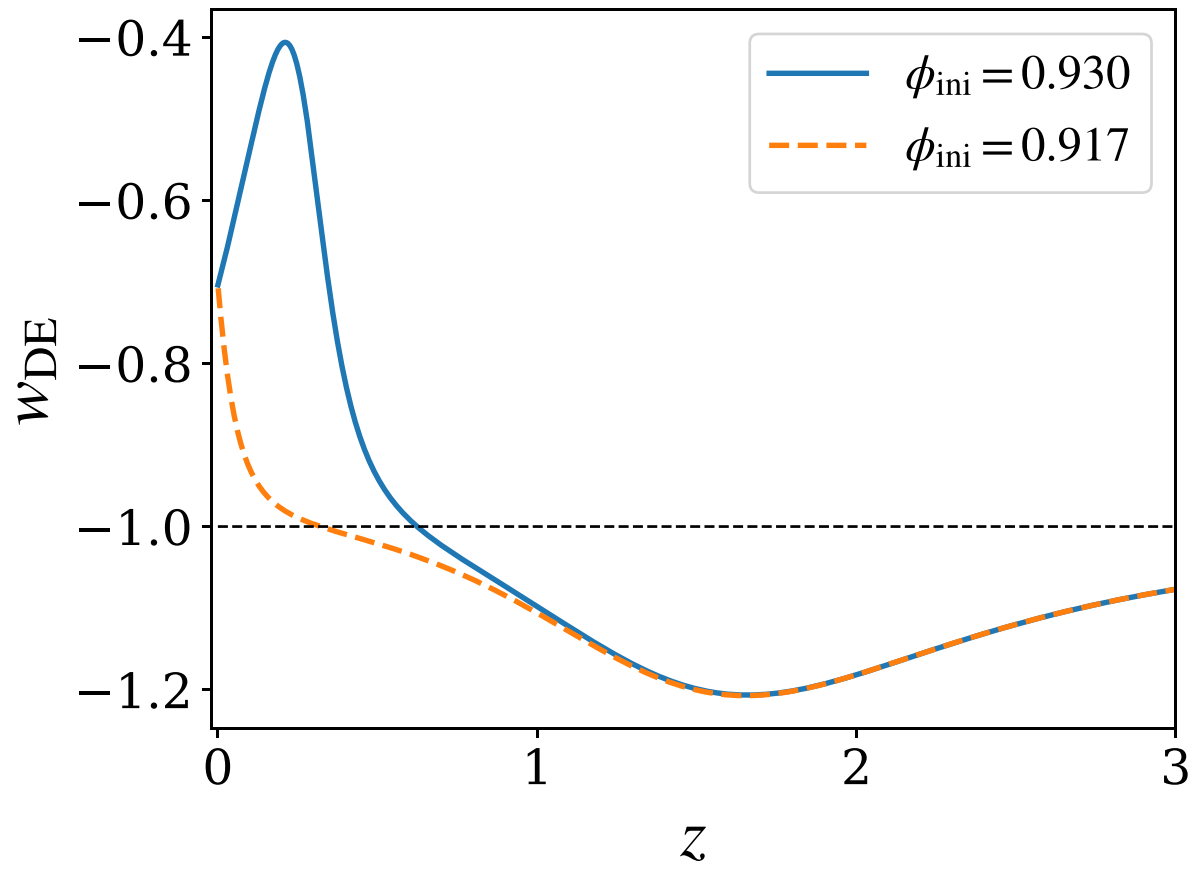

**Figure 6.** Late-time evolution of $w_{\rm DE}$ for $V_0 = 0.8 \times 10^{-8}\, m_{\rm P}^4$, $s = 25.8\, m_{\rm P}^{-1}$ and two particular values of $\phi_{\rm ini}$ that give the same present-day value of $w_{\rm DE,0}$. In the case with the greater $\phi_{\rm ini}$ value (solid curve), we see that $w_{\rm DE}$ crossed the phantom boundary earlier, and has peaked and is decreasing, tending to $-1$.

### 5.1 Background analysis

We solve the Friedmann equations for both a quintom and general $w_0 w_a$CDM model, fixing the same present-day best-fitting values obtained by DESI Collaboration (2025) for a $w_0 w_a$CDM cosmology, specifically with the DESI DR2 BAO+CMB+Pantheon+data set. In Figs 8 and 9 we plot our results for $w_{\rm DE}(z)$, the Hubble function $H(z)$ and the evolution of the matter and dark energy density parameters. For each observable we particularly focus on the redshift range that can be constrained by data.

We see that with a quintom model, we are able to obtain a late-time evolution of $w_{\rm DE}(z)$ fitting within $1\sigma$ of the best fit, although it exhibits a stronger oscillatory behaviour than the smooth transition modelled by the $w_0 w_a$CDM CPL parametrization. Interestingly, phantom crossing coincides at the same redshift. However, we note that if $w_{\rm DE}(z)$ were to be extrapolated to higher redshifts, a deviation starts to occur whereby the best-fitting $w_0 w_a$CDM parametrization favours a value of $w_{\rm DE}$ that continues to decrease, asymptoting to approximately $-1.35$ (the exact value depending on the data set being employed), while in the case of the quintom model $w_{\rm DE}$ would increase and always tend to $-1$, behaving almost like a cosmological constant.

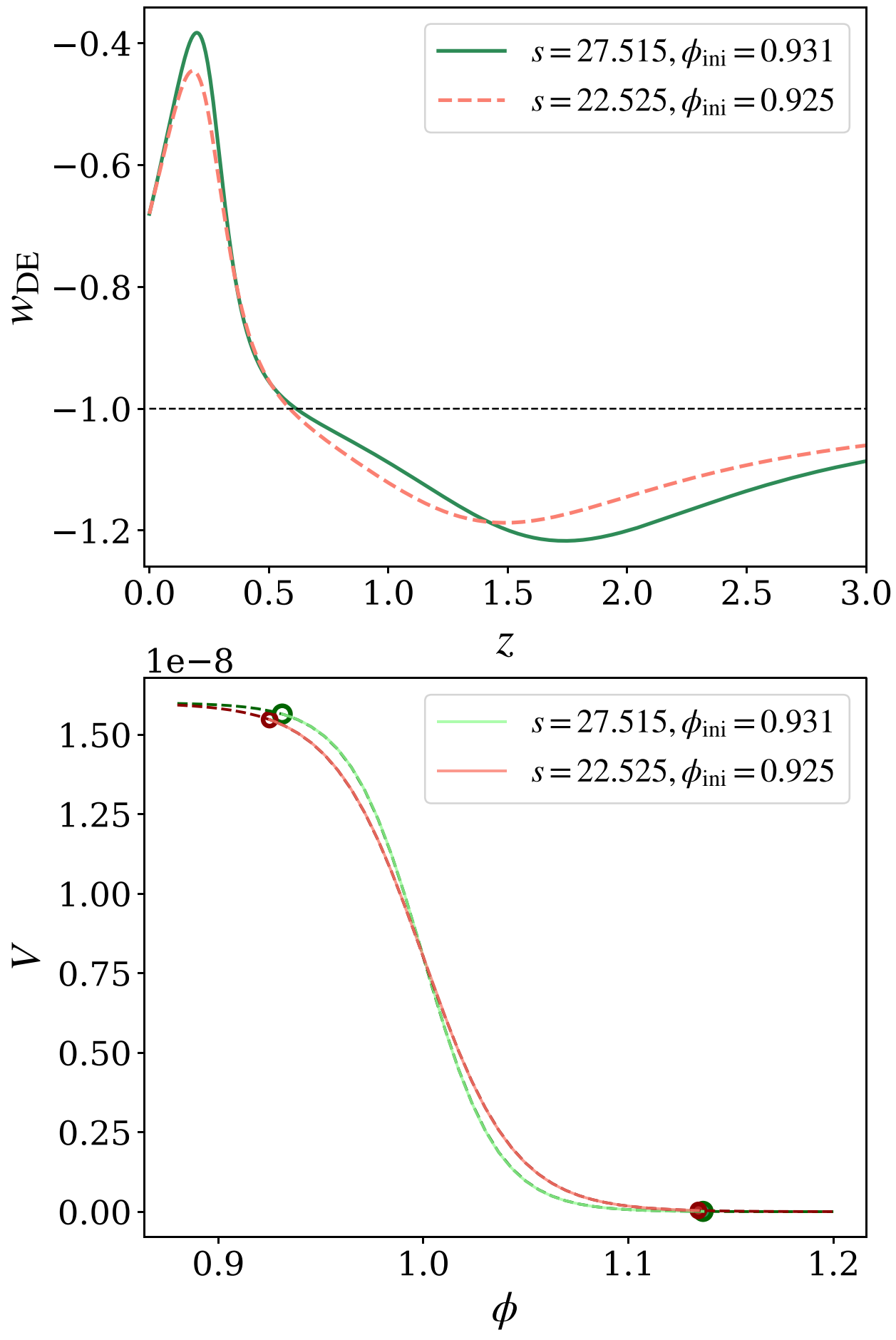

**Figure 7.** Top: plot of and $w_{\rm DE}(z)$ for $V_0 = 0.8 \times 10^{-8}\, m_{\rm P}^4$ and two sets of values for $s$ and $\phi_{\rm ini}$ (solid and dashed) which give the same $w_{\rm DE,0}$. The redshift of phantom crossing is approximately the same for both sets of values, while in the case of a smaller $\phi_{\rm ini}$ (dotted), the peak of $w_{\rm DE}$ is smaller (more negative). Bottom: corresponding plot of $V(\phi)$. The initial point of the field is marked out by the unfilled circles, which evolve to the point at the bottom of the potential marked out by the filled circles. We see that they evolve to roughly the same present-day value. An increase in $s$ (a sharper drop and thus flatter initial plateau) can be offset by allowing the $\phi$ field to start closer to the edge (making $\phi_{\rm ini}$ larger), giving the same initial value for the derivative of the field, and thus dynamics.

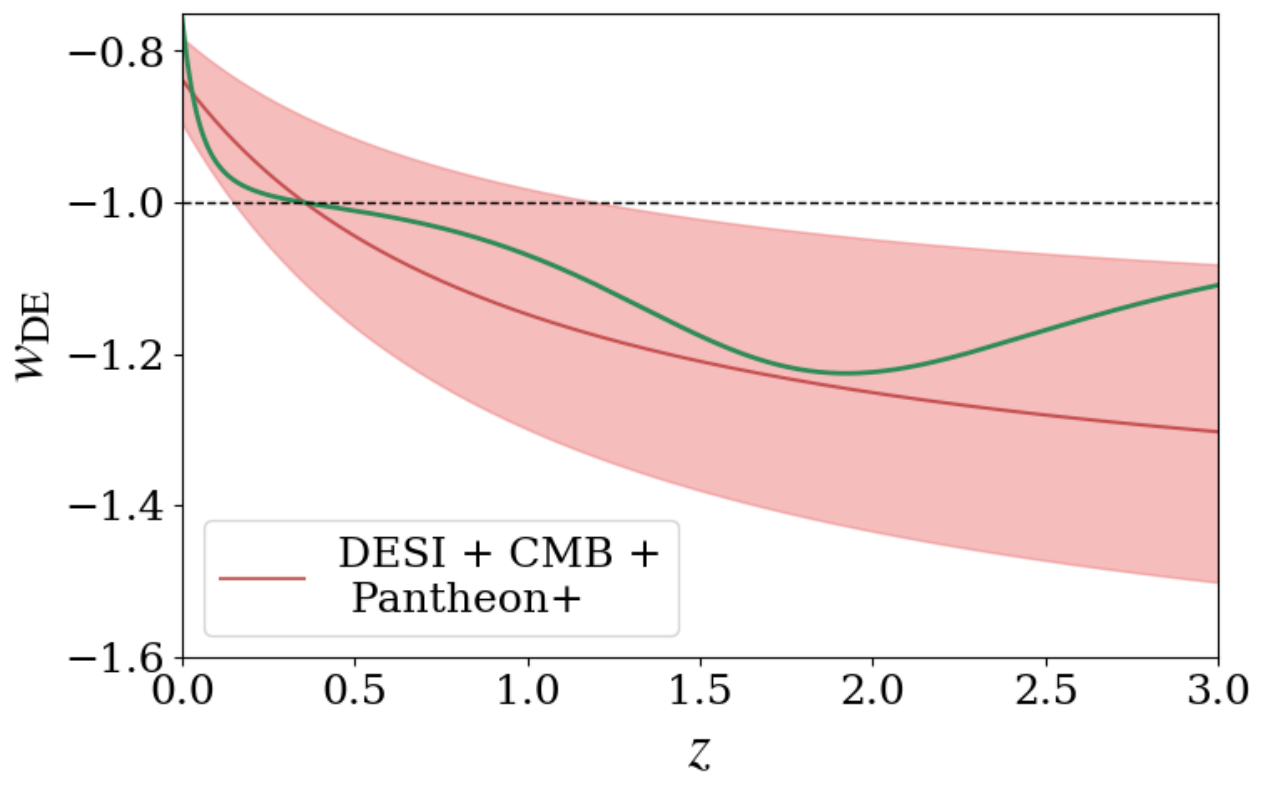

**Figure 8.** Plot of the late-time evolution of the dark energy EoS parameter, compared to the best fit obtained from DESI Collaboration (2025) using the DESI BAO+CMB+Pantheon+data set for a $w_0w_a$CDM cosmology. The shaded red region marks the $1\sigma$ uncertainty.

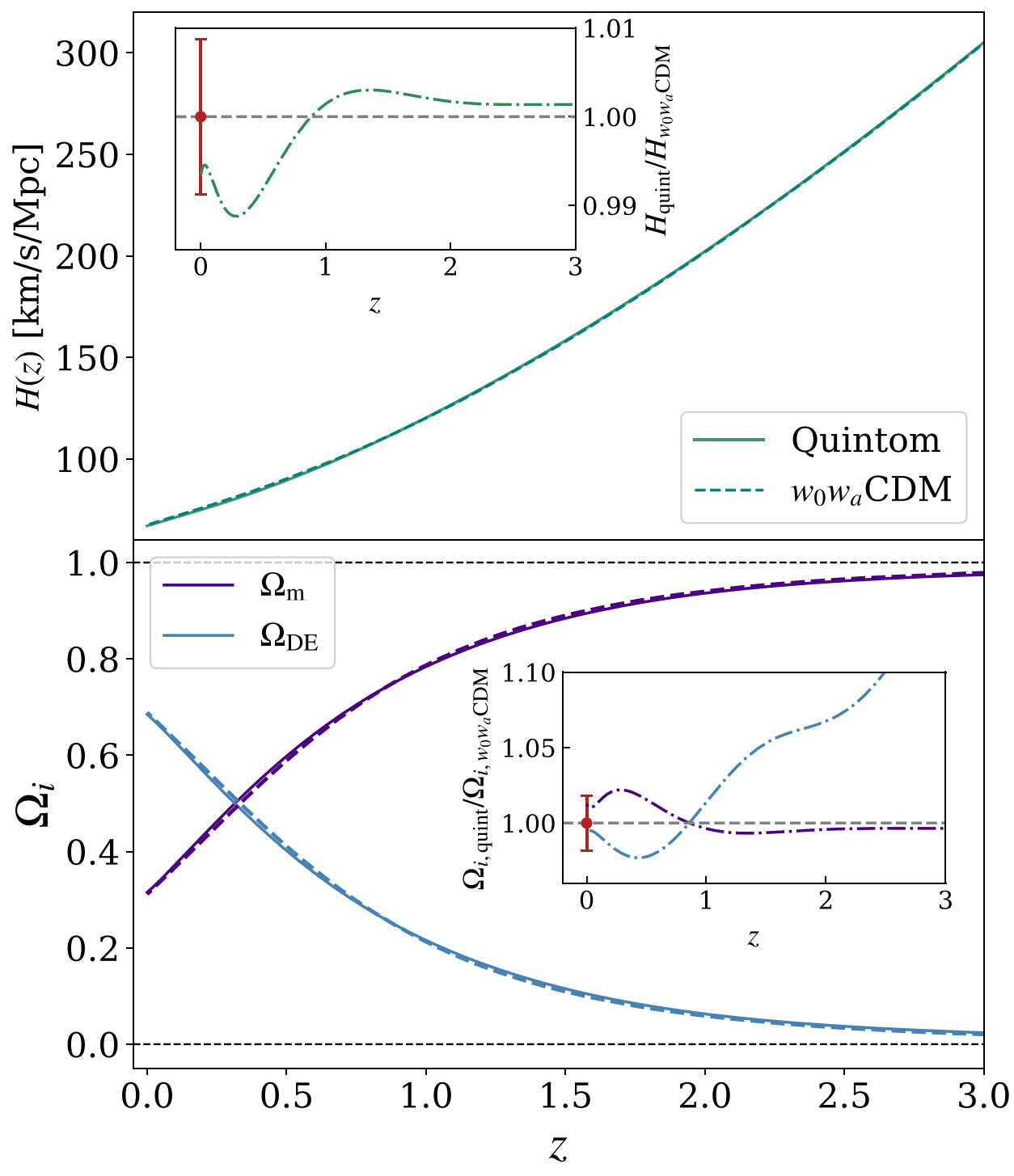

**Figure 9.** Top: plot of the Hubble function $H(z)$, for both the quintom model (solid) and the best fit $w_0w_a$CDM model (dashed). The inset plot shows the ratio of the Hubble function between the two models in dashed dotted lines. The red point with its associated $1\sigma$ error bar denotes the normalized best-fitting value of $H_0$ obtained with the DESI BAO+CMB+Pantheon+data set. Bottom: plot of the evolution of the dark matter density ($\Omega_{\rm m}$; purple) and the dark energy density ($\Omega_{\rm DE}$; blue) parameters for the quintom model (solid lines) and best-fitting$w_0w_a$CDM model (dashed lines). The inset plot shows the ratio of the $\Omega_i$'s between the two models in dashed dotted lines, with the red point and error bars marking the normalized best-fitting value and $1\sigma$ uncertainty of $\Omega_{\rm m}$ derived from the same DESI data set.

In terms of the Hubble function and $\Omega_{\rm m}$, we find that we can also obtain present-day values within $1\sigma$ constraints of the data. In Appendix A3 we present additional results where we selectively vary $\phi_{\rm ini}$ and $\psi_{\rm ini}$ in a quintom model. Comparing their evolutions, we note a deviation in both $H(z)$ and $\Omega_{\rm m}(z)$ at redshifts $0 < z < 1$ between the two models. This can be explained when looking at $\rho_{\rm DE}(z)$ for the $w_0w_a$CDM model, which evolves as $\rho_{{\rm DE},w_0w_a{\rm CDM}} \propto a^{-3(1+w_0+w_a)}e^{-3w_a(1-a)}$. At early times, $\rho_{{\rm DE},\,w_0w_a{\rm CDM}}$ increases monotonically, with $\rho_{{\rm DE},\,w_0w_a{\rm CDM}} < \rho_{\rm DE,\,quint}$, leading to a smaller $H(z)$ and larger $\Omega_{\rm m}$. At a redshift of $z \approx 0.8$, $V_{\rm quint}$ decreases due to the quintessence field taking dominance, and instead $\rho_{{\rm DE},\,w_0w_a{\rm CDM}} > \rho_{\rm DE,\,quint}$, where we then see a turnover of the ratio of $H(z)$ and $\Omega_{\rm m}$.

With a measurable difference in $H(z)$, this would invariably leave an imprint on the various cosmological distance relations, some of which can be directly constrained with BAO scale measurements. BAOs give the transverse and line-of-sight comoving distance relative to the sound horizon $r_d$, respectively given by

$$D_{\rm M}(z) = \frac{c}{H_0}\int_0^z \frac{{\rm d}z'}{H(z')/H_0}\,, \tag{31}$$

$$D_{\rm H}(z) = \frac{c}{H(z)}\,, \tag{32}$$

as well as the derived isotropic BAO distance $D_{\rm V}(z) = \left(z\,D_{\rm M}(z)^2 D_{\rm H}(z)\right)^{1/3}$. Here, the sound horizon is a function of the baryon and matter energy density scaled to *Planck* 2018 best-fit

$$r_d = 147.05\,{\rm Mpc} \times \left(\frac{\Omega_{\rm b}h^2}{0.02236}\right)^{-0.13}\left(\frac{\Omega_{\rm m}h^2}{0.1432}\right)^{-0.23}\left(\frac{N_{\rm eff}}{3.04}\right)^{-0.1}, \tag{33}$$

where $h \equiv H_0/100\ {\rm km\,s^{-1}\,Mpc^{-1}}$ and $N_{\rm eff}$ is the effective number of relativistic species, which we fix to be $N_{\rm eff} = 3.04$.

Given the Hubble function, we calculate the theoretical values of these three distances and compare them directly with the BAO measurements reported in table IV of DESI Collaboration (2025). We present these in Fig. 10, where we plot the ratio of these BAO distances to a fiducial cosmology, which, following the methodology of DESI Collaboration (2025), is taken to be the best-fitting *Planck* 2018 $\Lambda$CDM model. We also calculate the theoretical distances when adopting the best-fitting values of a $\Lambda$CDM model fitted to the DESI BAO+CMB data set (dashed blue) and a $w_0w_a$CDM model fitted to the DESI+CMB+Pantheon+data set (dashed dotted orange). We see that the theoretical evolution of the BAO distances in a quintom model roughly follow that of the $w_0w_a$CDM model, albeit with a weaker oscillatory behaviour whereby we do not see as large a decrease in $D_{\rm V}$, $D_{\rm M}$, and $D_{\rm H}$ at $z \approx 0.3$. This could be explained by the smaller value of $H(z)$ and its inverse relation to the comoving distances. However, it still provides a better fit to the data than a $\Lambda$CDM model. Of course, a full Markov chain Monte Carlo (MCMC) analysis would be required to robustly derive the best fit set of parameter values of our model, something which we shall leave for future work.

## 6 DISCUSSION

We have demonstrated how two-field quintom models with a hill-top or cliff-face potential can give rise to a dynamical dark energy model with a phantom to quintessence transition, as captured by the CPL $w_0w_a$ parametrization and hinted by recent DESI BAO data. It would hence be worth investigating if quintom models are distinguishable from a ΛCDM or $w_0w_a$CDM paradigm, in terms of the imprint they leave on the expansion history of the Universe and its growth of structure.

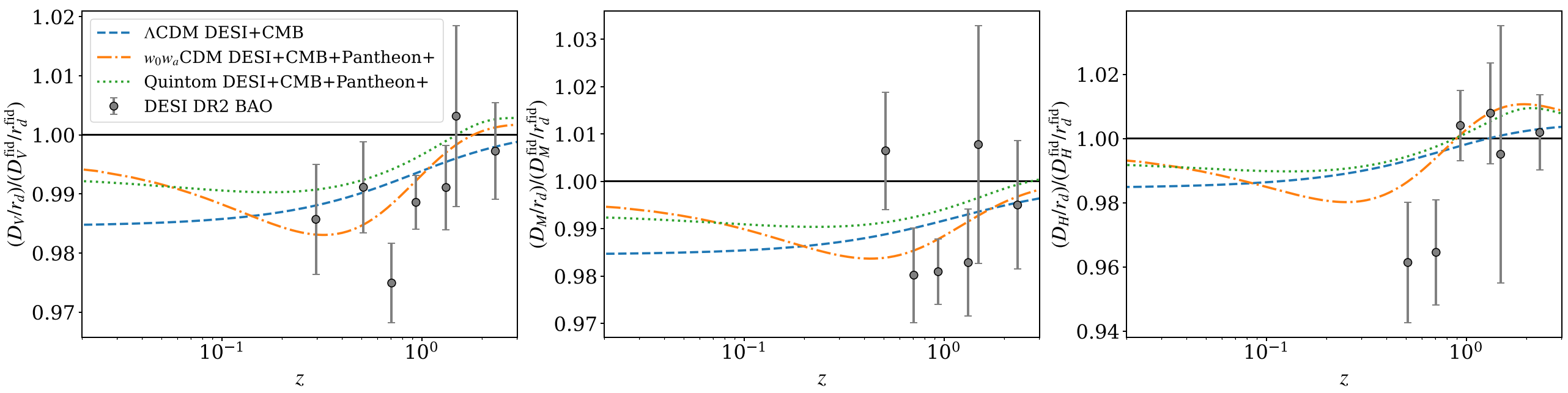

**Figure 10.** Ratio of the isotropic BAO distance to its fiducial value ($(D_V/r_d)/(D_V^{\rm fid}/r_d^{\rm fid})$; left), where we have adopted *Planck* 2018 ΛCDM best-fit (Planck Collaboration VI 2020) as the fiducial cosmology. We plot the theoretical ratios employing best-fitting values assuming ΛCDM cosmology with a DESI BAO+CMB data set (blue dashed), a $w_0w_a$CDM cosmology with a DESI BAO+CMB+Pantheon+ data set (orange dashed dot), and a quintom cosmology with a DESI BAO+CMB+Pantheon+data set (green dotted). We also plot the DESI BAO data points [the first seven measurements used for cosmology from table IV of DESI Collaboration (2025)] with their $1\sigma$ error bars in filled grey points. We do the same for the ratio of the transverse comoving distance [$(D_M/r_d)/(D_M^{\rm fid}/r_d^{\rm fid})$; middle] and line-of-sight comoving distance [$(D_H/r_d)/(D_H^{\rm fid}/r_d^{\rm fid})$; right].

In Section 5.1 we compared their background evolution, and found that within the low redshift range constrained by current data, a quintom model can provide a fairly accurate explanation of the cosmological paradigm currently preferred by the data. However, as mentioned, deviations in the dark energy EoS $w(z)$ start to manifest at higher redshifts, where we begin to see the fundamental differences between thawing scalar models and dynamical dark energy models: in our specific setup with the chosen initial values and hyperbolic tangent potential, the slow-roll regime presents a natural attractor solution during the epochs where dark energy is not the dominant component of the Universe. As such, the dark energy density $\rho_{\rm DE}$ remains constant and $w$ tends to $-1$. On the other hand, in dynamical $w_0w_a$CDM models, the dark energy density scales with a different relation and $w(z)$ does not possess attractor solutions. From this standpoint, measurements of $w(z)$ at higher redshifts would naturally be a good differentiator between these two particular models.

In Section 3.2, we discussed how the presence of dark energy perturbations affects the evolution of the gravitational potentials. R. R. Caldwell, R. Dave & P. J. Steinhardt (1998), J. Weller & A. M. Lewis (2003), G.-B. Zhao et al. (2005), and Y.-F. Cai et al. (2010) have investigated the effects of dark energy scalar field perturbations on physical observables such as the Cosmic Microwave Background (CMB) and the LSS. These arise mainly due to a change in $\dot\Phi$ and $\dot\Psi$ in the presence of an evolving dark energy EoS [also evident from equations (22) and (23)], which impacts the late-time Integrated Sachs–Wolfe (ISW) effect. When dark energy perturbations are taken into account, for an oscillating or quintom-A type model, this leads to an overall enhancement of the CMB temperature power spectrum at large scales, as well as of the matter power spectrum at all scales. During the phantom-dominated phase, $\rho_{\rm DE}$ and thus the dark energy perturbations are increasing, counteracting the late-time decay of the gravitational potentials and leading to an overall suppression of the ISW signal. The opposite happens for a quintessence-dominated scenario. How these opposing effects will play out for a quintom model when the two fields are combined then depends on the exact setup and chosen values of the initial conditions, as well as the nature of the field that dominates first. We shall leave this, along with a more in-depth investigation into the linear and non-linear matter perturbations of a quintom model, for future work.

Up to now, literature has placed constraints on quintom models using previous generation CMB data from WMAP, and SNe1a from the *Hubble Space Telescope* (B. Feng et al. 2005; G.-B. Zhao et al. 2005; J.-Q. Xia, B. Feng & X.-M. Zhang 2005; J.-F. Zhang & Y.-X. Gui 2010). It would thus be worthwhile to repeat this exercise with a quintom-B type model, and to confront it with more up-to-date data sets, cross-correlating CMB and LSS data to probe the ISW signal, as well as employing data from weak gravitational lensing (WL), which will be able to probe the evolution of dark energy to markedly non-linear scales (S. Joudaki, A. Cooray & D. E. Holz 2009).

## 7 CONCLUSION

In this work, we have studied the two-scalar field quintom model, whereby a quintessence-like and phantom-like field work to achieve phantom crossing behaviour that is stable to gravity and perturbations. We have particularly focused on quintom-B type models, where the transition evolves from a phantom to quintessence regime with $w < -1$ to $w > -1$, as has been favoured by recent BAO data from DESI DR 2. While such a transition has historically been proven to be more difficult to realize, we have constructed a physically motivated framework featuring naturally arising hill-top or cliff-face potentials that can achieve this while requiring minimal degrees of freedom. In particular, we considered a scenario in which both fields roll in opposing directions along a shared potential, and studied two specific forms for the potential: a Gaussian and a hyperbolic tangent. By varying the initial conditions, we find a natural attractor solution at early times whereby the system tends to $w = -1$, mimicking ΛCDM behaviour, before the effects of the field begin to dominate, and we see a change in the evolution of $w(z)$ at late times. We observe that the evolution of the system is sensitive to the initial values of the fields, thus admitting a rather restricted parameter space given the precision of observational data already at hand that can constrain the model. This would hence require more robust phenomenological motivations to explain such a 'why now' conundrum, since no explicit attractor solution appears to arise in this regard.

When comparing our results to the DESI BAO data, we see that we are able to reproduce the behaviour of $w(z)$ derived from the best-fitting values of the data when adopting a $w_0w_a$CDM CPL parametrization. We further compare other background quantities, such as the Hubble function, the evolution of the matter density parameter, as well as the BAO distances, finding once again a reasonable agreement between the values derived from a quintom model and the data. However, we do note that this is solely a first-order approximation, and a full MCMC analysis with the data would be required to derive robust constraints on our model. We shall leave this as forthcoming work. Finally, we discussed potential ways to constrain such a model in terms of the imprints it might leave on $w(z)$ at high redshifts, as well as a suppression of the matter power spectrum and the late-time ISW signal of the CMB when dark energy perturbations are taken into account. Such avenues of exploration would also be well worth deeper investigation.

We find ourselves in an exciting time of cosmological research, with current data potentially unveiling new physics, and more incoming at unprecedented levels of precision, such as LSS data from *Euclid* (Euclid Collaboration 2025) and LSST (Ž. Ivezić et al. 2019), as well as CMB data from the Simons Observatory (N. Galitzki et al. 2024) and Litebird (M. Hazumi et al. 2020). In this context, constructing well-motivated theoretical frameworks – such as the quintom model explored here – is crucial. These efforts will not only help interpret the growing wealth of cosmological data but may also guide us toward a deeper understanding of the fundamental nature of dark energy and of our Universe.

## ACKNOWLEDGEMENTS

LWKG thanks the University of Edinburgh School of Physics and Astronomy for a postdoctoral Fellowship, while ANT thanks the UK Science and Technology Funding Council (STFC) for support. For the purpose of open access, the authors have applied a Creative Commons Attribution (CC BY) licence to any Author Accepted Manuscript version arising from this submission.

## DATA AVAILABILITY

The data underlying this article will be shared on reasonable request to the corresponding author.

## APPENDIX A: VARYING INITIAL CONDITIONS

### A1 Gaussian potential

In Figs A1 to A4, we present the full results of our investigation on the impact of varying the initial conditions of the field and the potential: $\{\dot{\phi}_{\rm ini}, \dot{\psi}_{\rm ini}, \phi_{\rm ini}, \psi_{\rm ini}, V_0, \mu, \sigma\}$, for a Gaussian potential given by equation (27).

First, we see that varying the initial speed of either field does not alter the late time dynamics of the system, as can be seen from the first two subplots of the four figures. Regardless of the values of $\dot{\phi}_{\rm ini}$ and $\dot{\psi}_{\rm ini}$, the system tends to a slow-roll solution during the radiation dominated era, with $\dot{\phi} \approx \dot{\psi} \approx 0$. The fields then evolve similarly when they are unfrozen at late times.

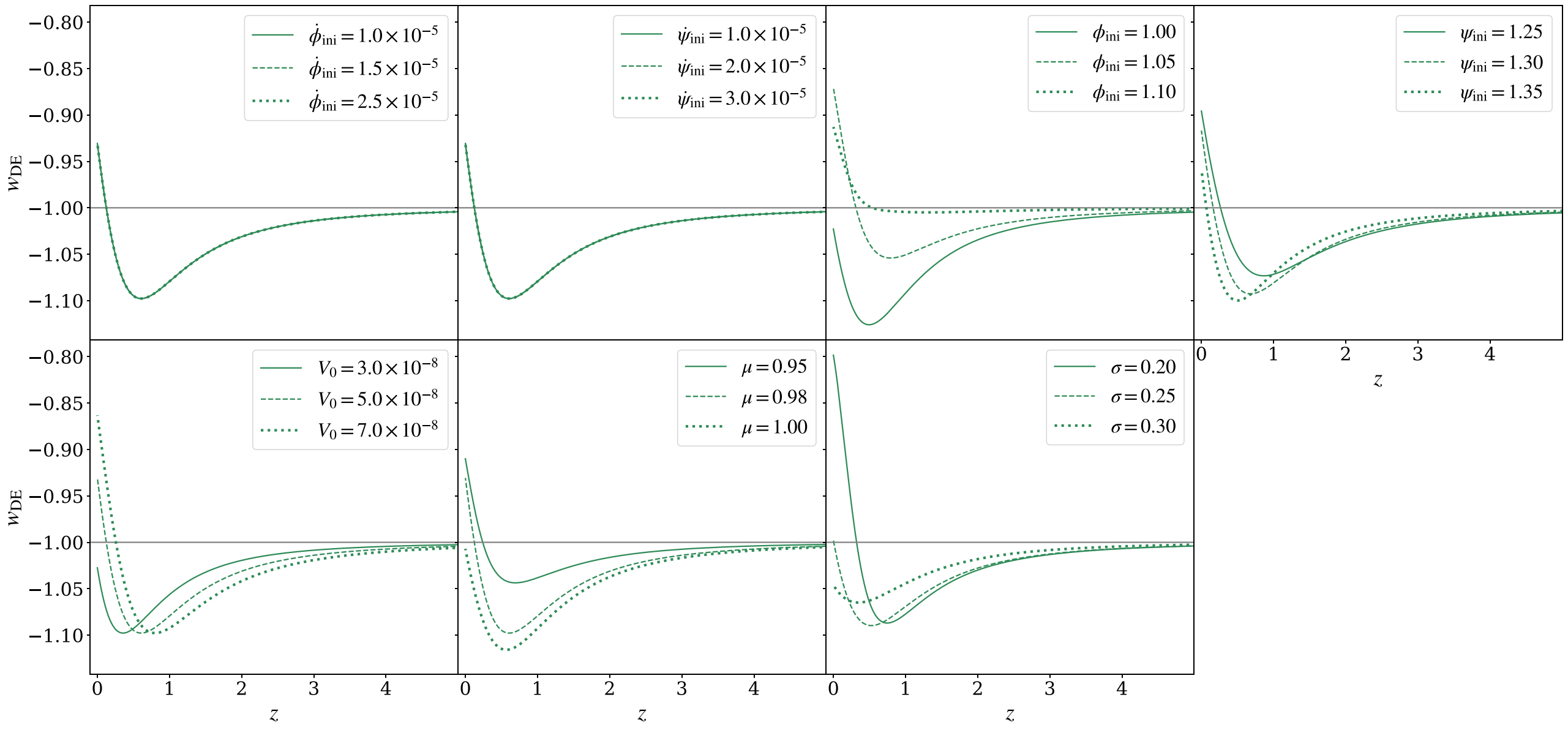

**Figure A1.** Plots of the evolution of the dark energy EoS parameter $w_{\rm DE}(z)$, when varying the seven parameters (left to right, top to bottom): $\{\dot{\phi}_{\rm ini}, \dot{\psi}_{\rm ini}, \phi_{\rm ini}, \psi_{\rm ini}, V_0, \mu, \sigma\}$ for a Gaussian potential. We choose three values to vary each parameter by, and present each case in solid, dashed, and dotted lines. The $w = -1$ boundary is drawn out in a solid line.

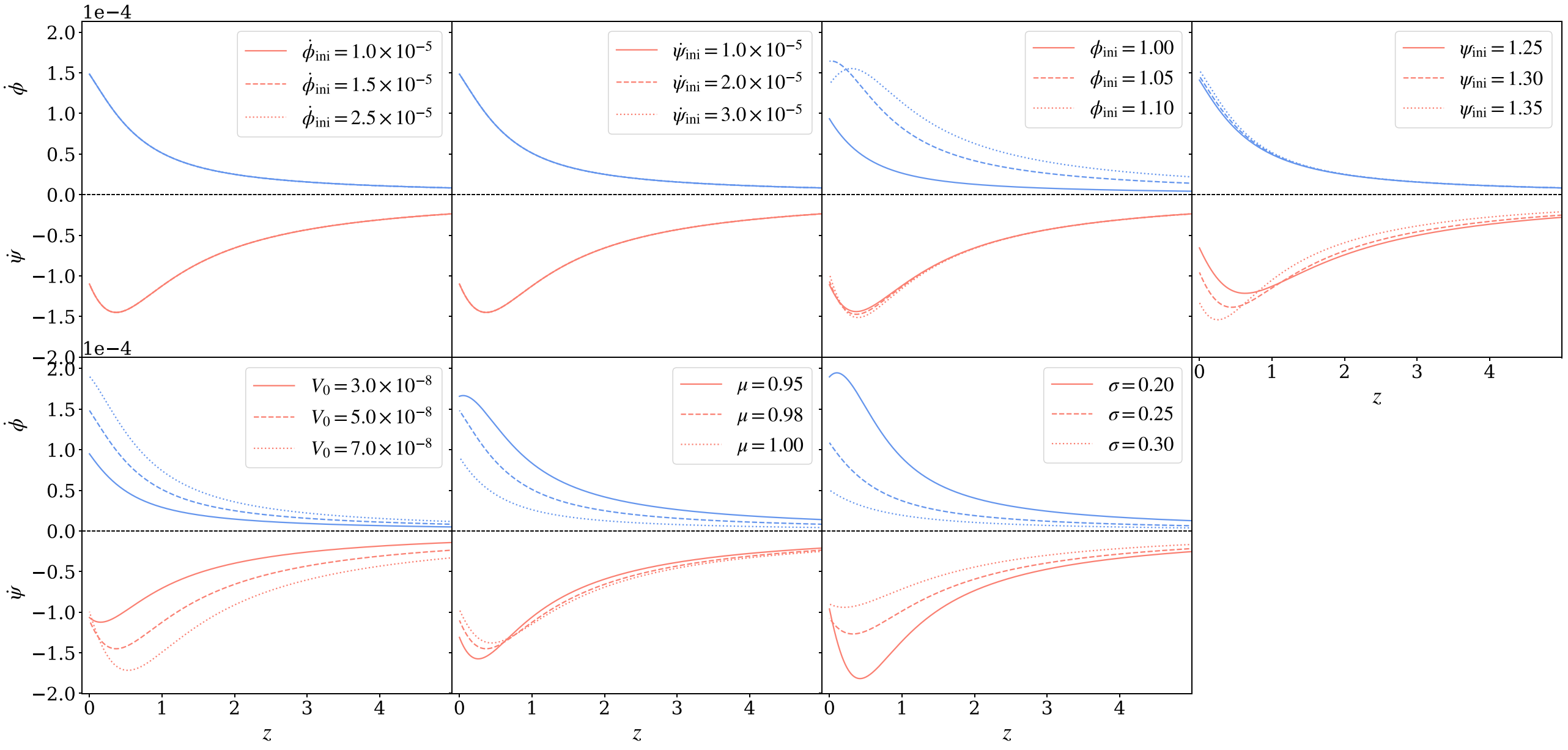

**Figure A2.** Plots of the speeds of the phantom $\psi$ field (pink) and quintessence $\phi$ field (blue) as a function of redshift, varying the same set of parameters over the same set of values for a Gaussian potential. The $\dot{\psi} = 0$ line has been drawn out in dashed lines.

Next, by starting the quintessence field further up the potential, we see that this decreases $\phi_{\rm ini}$ and, in turn, the speed at which it rolls down the potential when the field eventually thaws. When $\phi_{\rm ini}$ is too far up the potential (for example, when $\phi_{\rm ini} = 1.00\, m_{\rm P}$ in Fig. A4), $\partial V/\partial\phi$ is too small for the $\phi$ field to gain enough speed rolling down the potential, such that $\dot{\phi} < \dot{\psi}$ and the system is always phantom-dominated. Conversely, starting the phantom field further down the potential (increasing $\psi_{\rm ini}$) increases $\partial V/\partial\psi$ and thus the speed at which it travels up the potential. This manifests as a sharper decrease in $w(z)$ before the transition.

Varying $V_0$, $\mu$, and $\sigma$ affect the behaviour of both fields: the overarching trend is that increasing $V_0$ (or decreasing $\mu$) increases $\partial V/\partial\phi$ increases the speeds of both fields, which can once again be inferred from equation (28). We then generally see the inverse trend for the variation of $\sigma$, whereby a narrower potential would naturally lead to a sharper decrease (increase) in the $\phi$ ($\psi$) field. In the scenario of $\sigma = 0.20\, m_{\rm P}$, we see that the $\psi$ field gains enough speed that it evolves past the peak (denoted by the vertical grey line) and begins to roll down the other side of the potential. Since the dynamics of a phantom field can be thought of as the inverse of a quintessence field, the peak would thus be a minima, and hence the phantom field would oscillate around it and eventually come to rest at the highest point, presenting a natural attractor solution. With a more gradual slope (larger $\sigma$), this would give a smaller $\dot{\phi}$, leading to either phantom crossing occurring at lower redshifts with a smaller value of $w_{\rm DE,0}$, or no crossing at all.

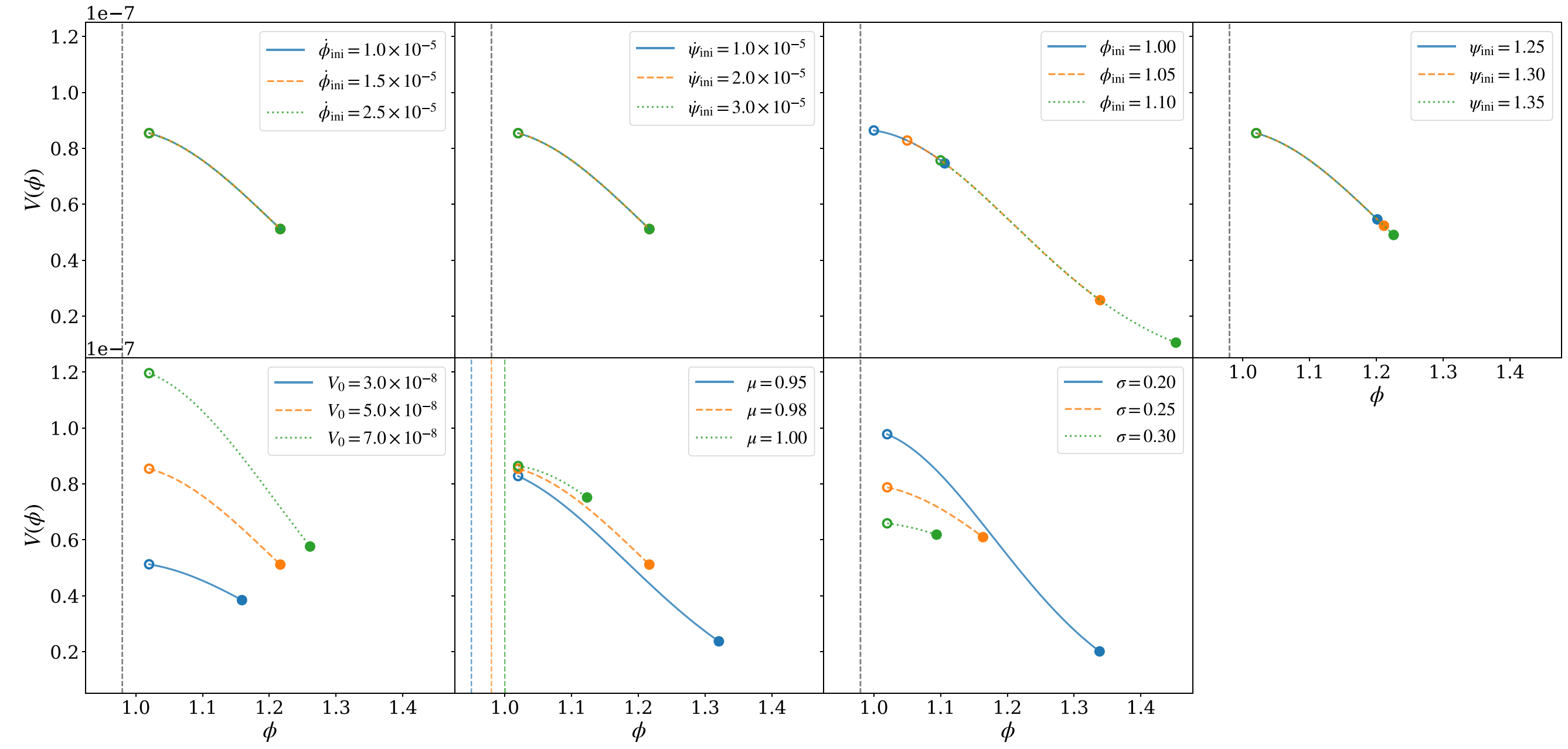

**Figure A3.** Plot of the potential of the quintessence field $V(\phi)$ as a function of $\phi$, varying the same set of parameters over the same set of values for a Gaussian potential. The start and end points of the field's position are marked by unfilled and filled circles, respectively. In this case, the field is rolling down the potential (increasing in $\phi$). The peak of the potential (i.e. the value of $\mu$) is marked as a vertical dashed line, or as the respective coloured lines in the case where $\mu$ is varied.

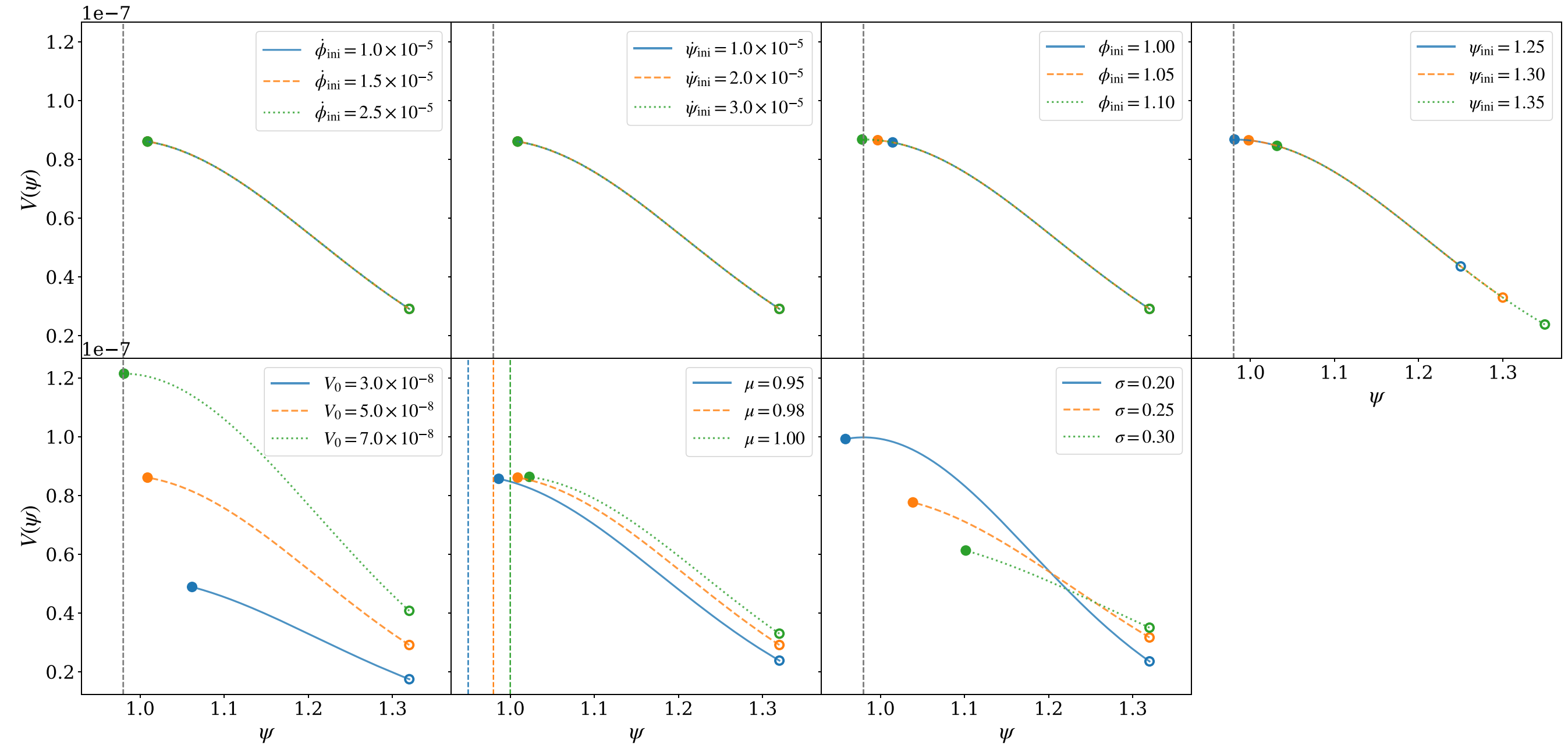

**Figure A4.** Similar plot as above, but for the phantom $\psi$ field. In this case, the field is rolling up the potential (decreasing in $\psi$).

### A2 Hyperbolic tangent potential

We present in Figs A5to A8, the full results of our investigation into the effects of varying the different initial parameters of the system: $\{\dot{\phi}_{\rm ini}, \dot{\psi}_{\rm ini}, \phi_{\rm ini}, \psi_{\rm ini}, V_0, s\}$, with a hyperbolic tangent potential given by equation (29).

We see generally similar trends to those of a Gaussian potential, whereby varying the initial speeds of the fields has minimal impact on the evolution of the system. On the other hand, in terms of variations of the initial field values, the effects on their dynamics are decoupled at the redshift where phantom crossing occurs: changes in $\phi_{\rm ini}$ dictate the resultant value of $w_{\rm DE,0}$ while $\psi_{\rm ini}$ predominantly changes the magnitude of the decrease in $w_{\rm DE}$ before phantom crossing. Finally, we see that the quintessence field is most sensitive to changes in the parameters of the potential, $V_0$ and $s$: for a greater amplitude of $V$, the $\phi$ field evolves to different present-day values, with $\psi_0$ remaining approximately constant. This then results in a larger (i.e. less negative) value of $w_{DE,0}$, while the redshift at which the phantom crossing occurs is approximately the same. When varying the slope of the potential through $s$, a smaller value of $s$ leads to larger $w_{DE}$ as the quintessence field is able to roll further down the potential.

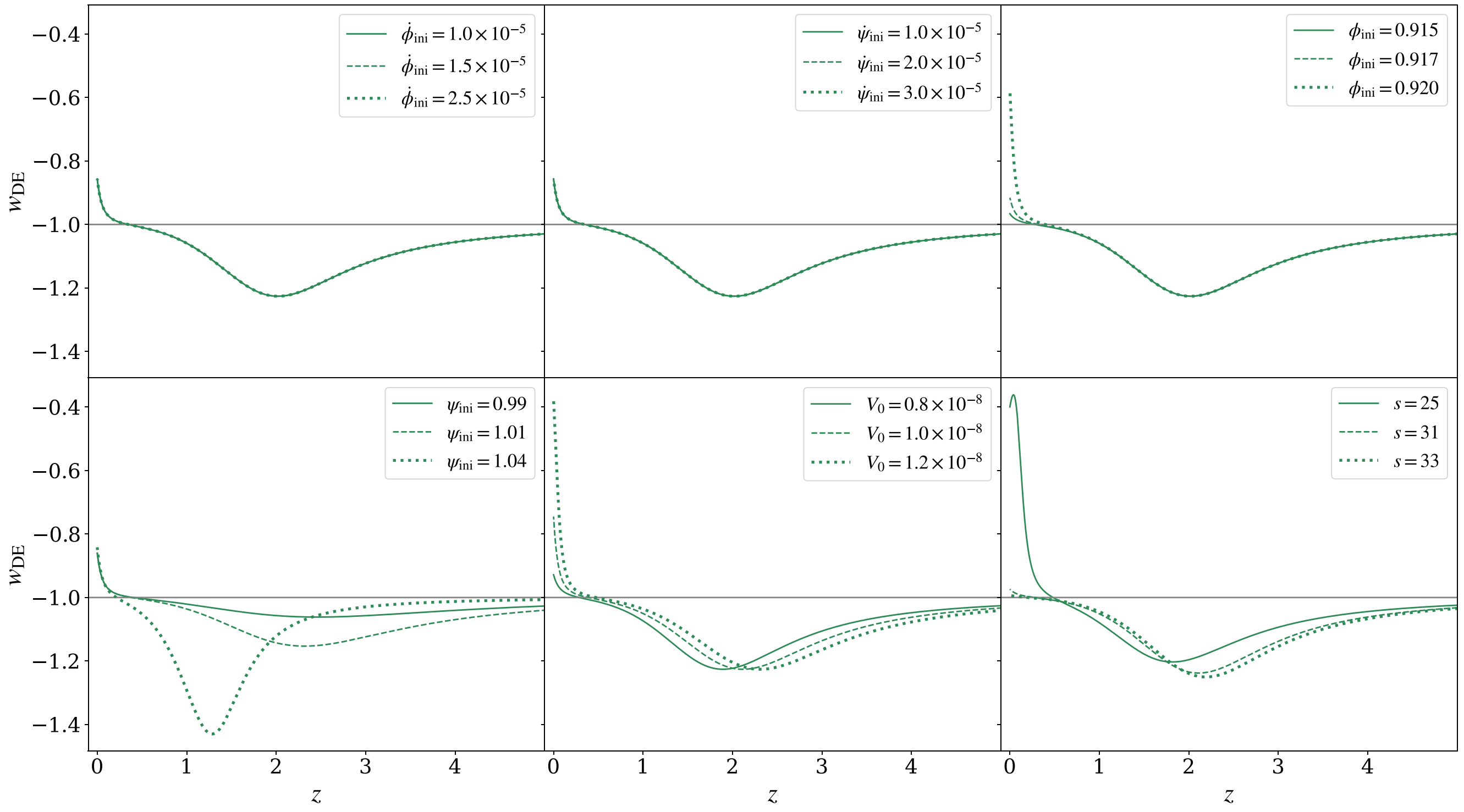

**Figure A5.** Plot of the dark energy EoS parameter as a function of redshift, when varying the parameters (top to bottom, left to right): $\{\dot\phi_{\rm ini}, \dot\psi_{\rm ini}, \phi_{\rm ini}, \psi_{\rm ini}, V_0, s\}$ for a hyperbolic tangent potential. We vary each parameter by three values, and plot the different $w_{\rm DE}(z)$ in solid, dashed and dotted lines. The $w = -1$ boundary has been drawn out with a grey line.

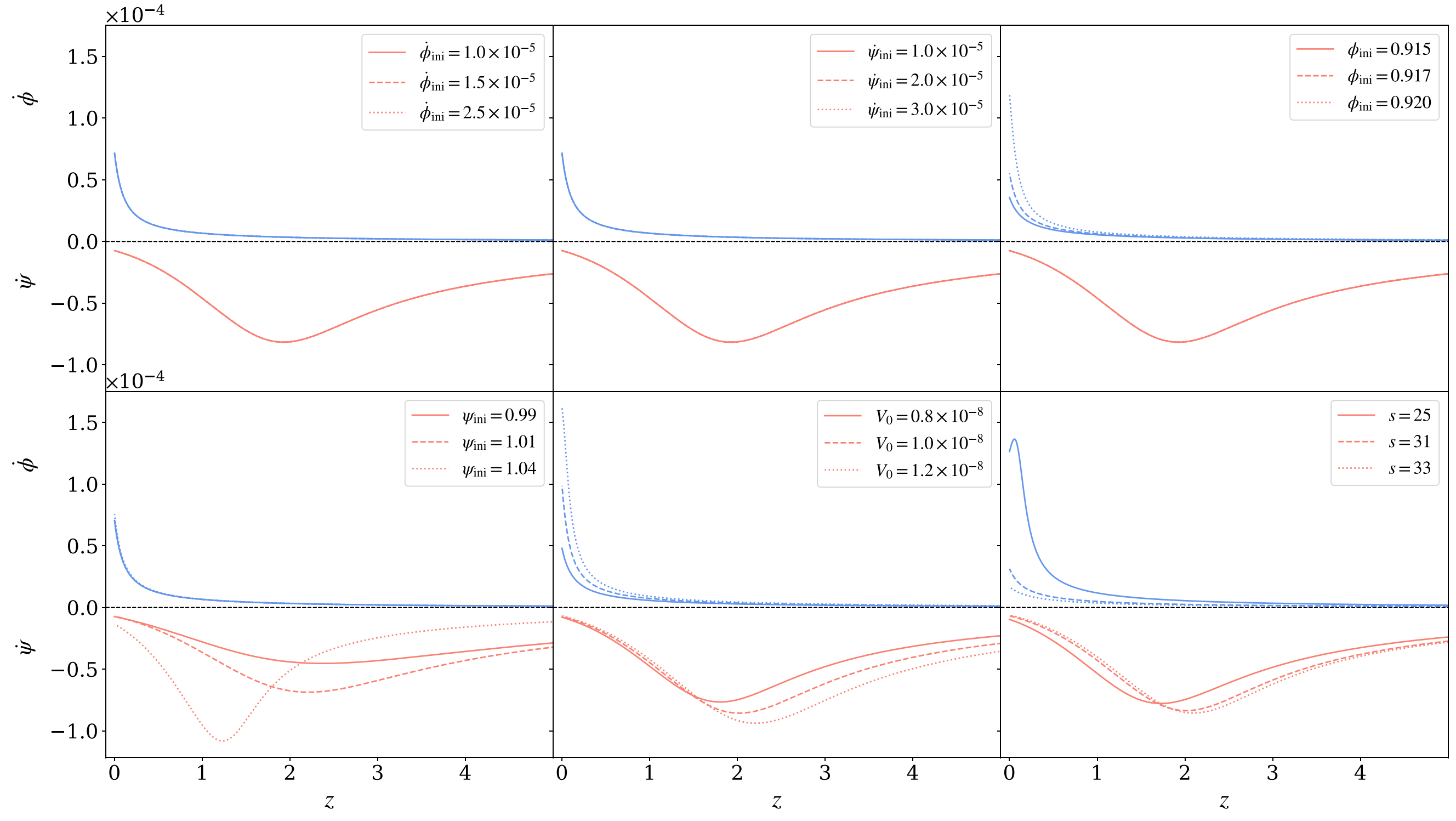

**Figure A6.** Plots of the speeds of the phantom $\psi$ field (pink) and quintessence $\phi$ field (blue) as a function of redshift, varying the same set of parameters over the same set of values for a hyperbolic tangent potential. The $\dot\psi = 0$ line has been drawn out in dashed lines.

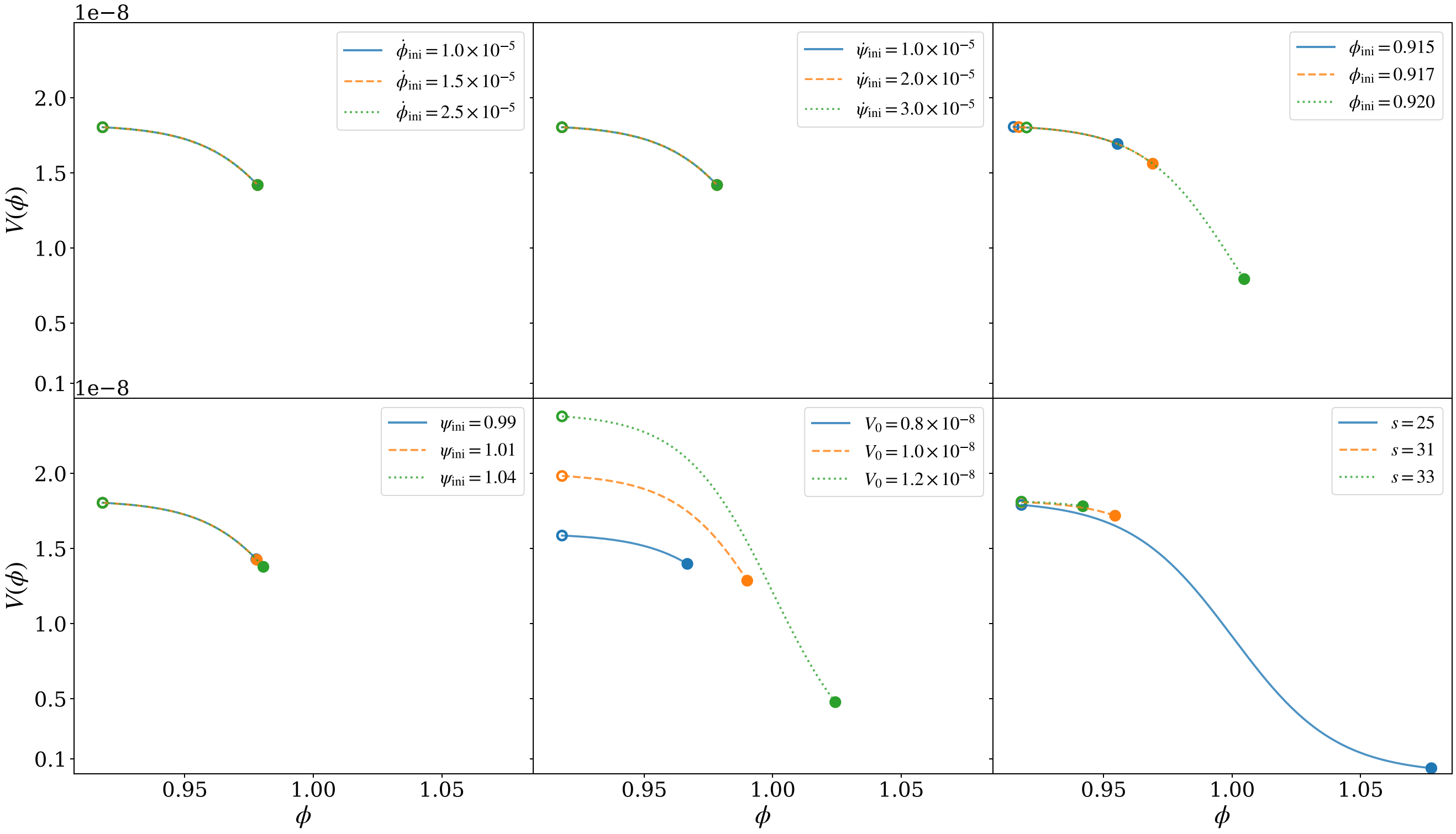

**Figure A7.** Plot of the potential of the quintessence field $V(\phi)$ as a function of $\phi$, varying the same set of parameters over the same set of values for a hyperbolic tangent potential. The start and end points of the field's position are marked by unfilled and filled circles, respectively. In this case, the field is rolling down the potential (increasing in $\phi$).

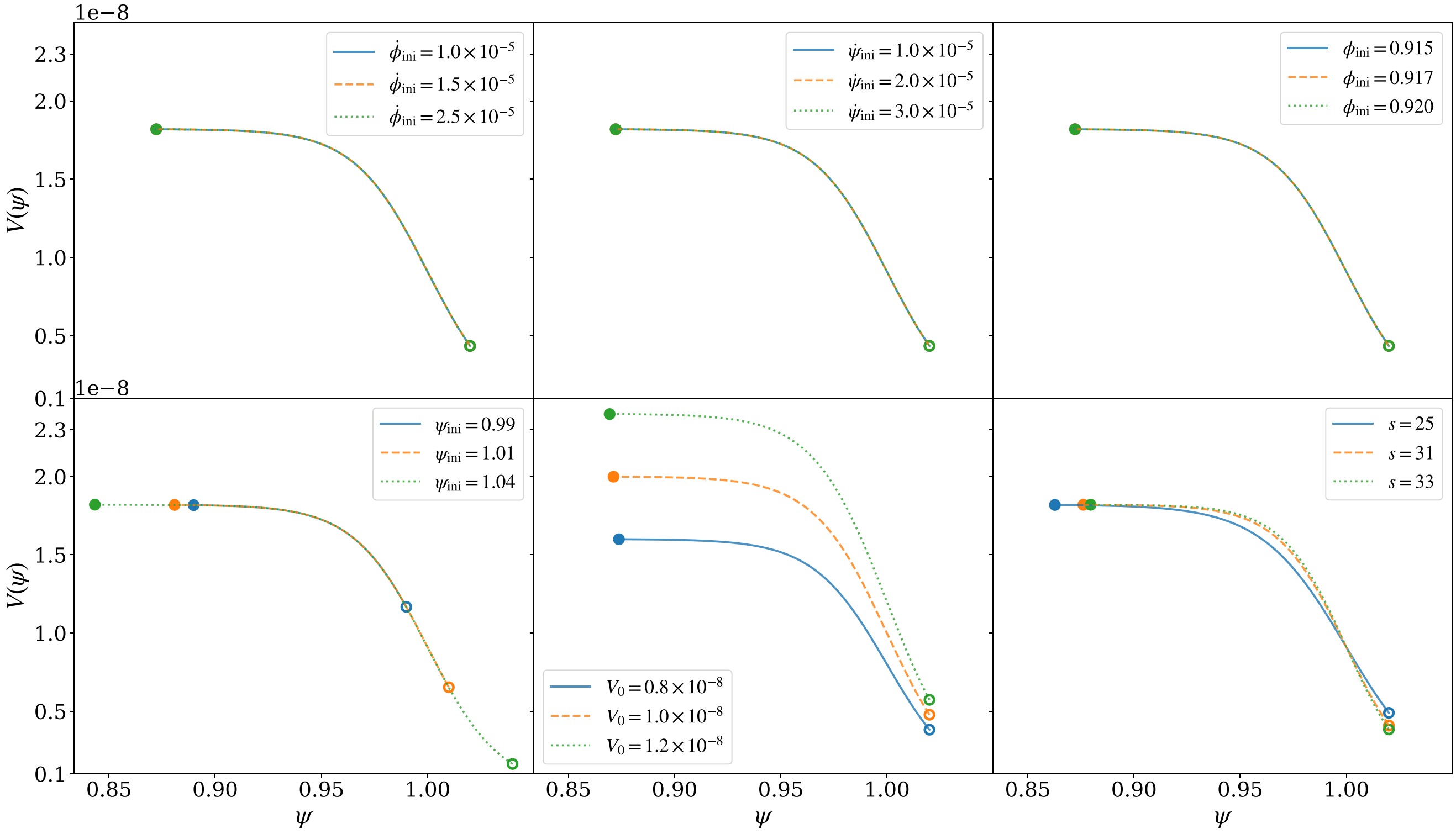

**Figure A8.** Similar plot as above, but for the phantom $\psi$ field. In this case, the field is rolling up the potential (decreasing in $\psi$).

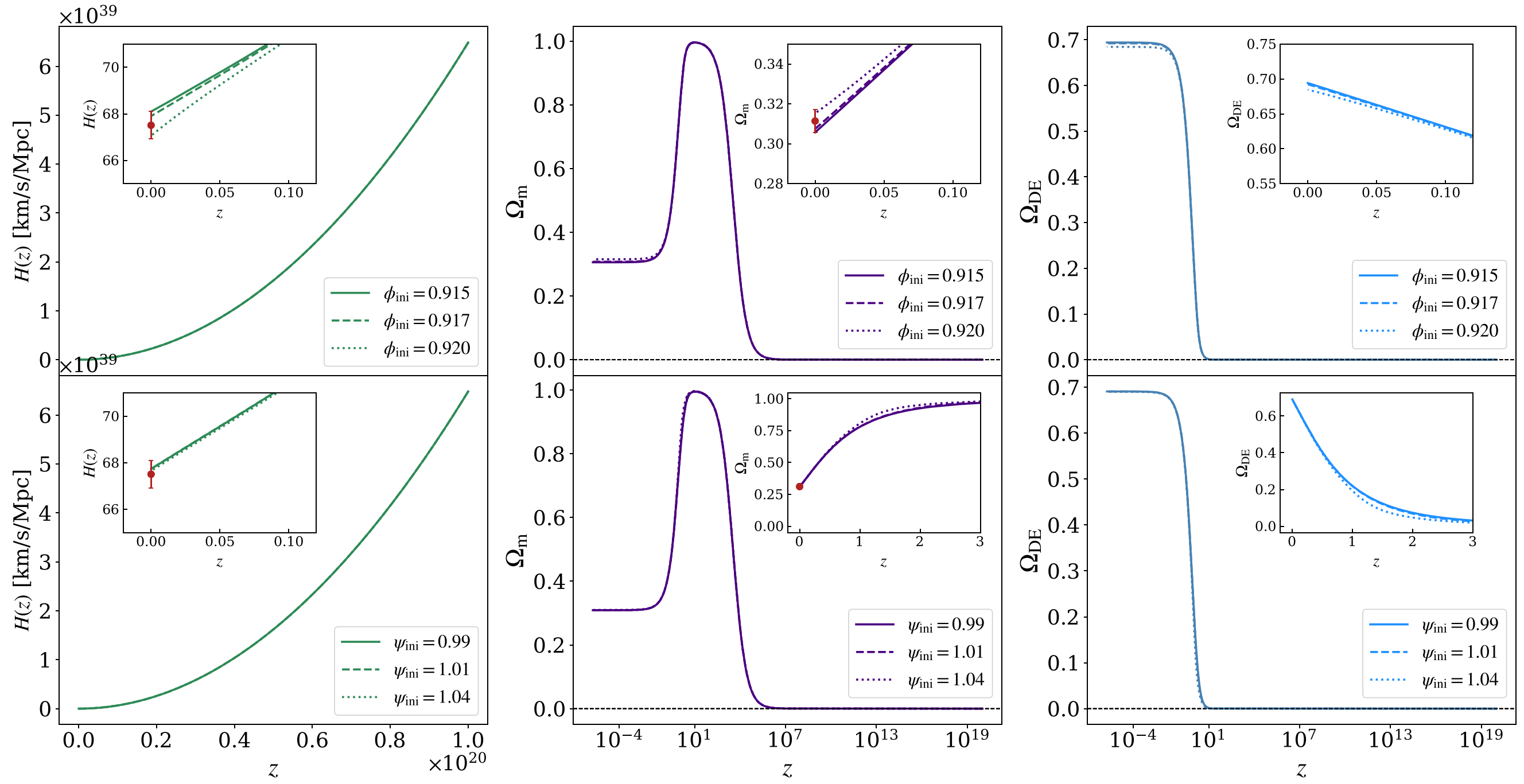

**Figure A9.** Top row: Plot of the Hubble function (left), $\Omega_m$ (middle), and $\Omega_{DE}$ (right) as a function of $z$ when varying $\phi_{ini}$. The inset plot shows a zoom-in at late times ($0 < z < 0.12$), where we include the best-fit and $1\sigma$ error bar for $H_0$ and $\Omega_{m,0}$ with the DESI BAO+CMB+Pantheon+data set for a $w_0w_a$CDM cosmology. Bottom row: Similar to the first row, instead varying $\psi_{ini}$. In the bottom row, the inset plots of $\Omega_m$ and $\Omega_{DE}$ have been set to redshifts $0 < z < 3$ to highlight the changes in the physical quantities that occur at slightly higher redshifts when varying $\psi_{ini}$.

### A3 Impact on physical quantities

We present in Fig. A9 the Hubble function and evolution of the energy density parameters when varying $\phi_{ini}$ and $\psi_{ini}$ for a hyperbolic tangent potential. Generally we see that varying $\psi_{ini}$ does not impact its present-day values ($H_0$ and $\Omega_{i,0}$) as much as $\phi_{ini}$. Rather, it works to change $\Omega_m$ and $\Omega_{DE}$ at slightly higher redshifts of approximately $1 < z < 2$. This is expected since variations in $\phi_{ini}$ affect the expansion history at low redshifts after phantom crossing has occurred, while $\psi_{ini}$ affects its dynamics before that. Increasing $\phi_{ini}$, as we have seen in Fig. A7, increases the extent to which it rolls down the potential, decreasing $V$ and hence the total dark energy density $\rho_{DE}$. This works to decrease the Hubble function and $\Omega_{DE} \equiv \rho_{DE}/H^2$ overall. On the other hand, since the dark matter and baryon sectors remain unchanged in a quintom model, $\Omega_m$ then increases with decreasing $H$. We see a similar effect when increasing $\psi_{ini}$, although this occurs at earlier times.

This paper has been typeset from a TEX/LATEX file prepared by the author.